%% file: main.tex
\documentclass[prl,aps,twocolumn,preprintnumbers,amssymb,nobibnotes,nofootinbib,raggedbottom,10pt]{revtex4-1}
\usepackage{amsmath}
\usepackage{hyperref,graphicx,array,multirow,color,amsmath,mathrsfs,titlesec,shuffle,tikz}
\usepackage{mathtools}
\input{header_data.tex}

\newcommand{\Ddots}{\hbox to 1em{.\hss.\hss.\hss}}

\begin{document}
\title{An All-loop Soft Theorem for Pions}
\author{Christoph Bartsch$^\spadesuit$}
\author{Karol Kampf$\,^\spadesuit$}
\author{Ji\v r\' i Novotn\' y$^\spadesuit$}
\author{Jaroslav Trnka$^{\spadesuit\Diamond}$}

\affiliation{$^\spadesuit$Institute for Particle and Nuclear Physics, Charles University, Prague, Czech Republic}
\affiliation{$^\Diamond$Center for Quantum Mathematics and Physics (QMAP), University of California, Davis, CA, USA}

\begin{abstract} 
In this letter, we discuss a generalization of the Adler zero to loop integrands in the planar limit of the $SU(N)$ non-linear sigma model (NLSM). While possible to maintain at one-loop, the Adler zero for integrands is violated starting at the two-loop order and is only recovered after integration. Here we propose a non-zero soft theorem satisfied by loop integrands with any number of loops and legs. This requires a generalization of NLSM integrands to an off-shell framework with certain deformed kinematics. Defining an {\it algebraic soft limit}, we identify a particularly simple non-vanishing soft behavior of integrands, which we call the \textit{algebraic soft theorem}. We find that the proposed soft theorem is satisfied by the `surface' integrand of Arkani-Hamed, Cao, Dong, Figueiredo and He, which is obtained from the shifted ${\rm Tr}\phi^3$ surfacehedron integrand. Finally, we derive an on-shell version of the algebraic soft theorem that takes an interesting form in terms of propagator renormalization factors and lower-loop integrands in a mixed theory of pions and scalars. 
\end{abstract}

\maketitle

\vspace{-0.3cm}

\section{Introduction}

\vspace{-0.3cm}

The Adler zero is a fundamental property of scattering amplitudes in theories with shift symmetry. A prime example of such a theory is the $SU(N)$ non-linear sigma model (NLSM), which describes the low-energy sector of QCD. As such, both tree- and loop-level amplitudes in the NLSM vanish when any momentum is taken soft. We start by writing the $n$-point tree amplitude ${\cal A}_n^{\rm tree}$ as a sum of \emph{flavor-ordered} amplitudes \cite{Kampf:2012fn,Kampf:2013vha},
\begin{align}
\hspace{-0.2cm} {\cal A}_n^{\rm tree} = \sum_{\sigma} {\rm Tr}(T^{a_1}T^{\sigma(a_2)}\dots T^{\sigma(a_n)})A_n\left(1\sigma(23\dots n)\right), \label{decomp}
\end{align}
where $\sigma$ runs over all permutations of labels $\lbrace 2,\ldots,n\rbrace$. The ordered amplitude $A_n(123\dots n)$ has only poles and factorization channels consistent with a canonical ordering of labels. Not only does the full amplitude ${\cal A}_n$ enjoy the Adler zero, but also the ordered amplitudes satisfy
\begin{equation}
\lim_{p_i\rightarrow 0} A_n = 0. \label{soft0}
\end{equation}
At the tree level, the above \emph{soft theorem}, together with consistent factorization, fix NLSM amplitudes uniquely. This fact was leveraged to formulate on-shell recursion relations for the tree-level S-matrix and lead to many other important insights \cite{Cheung:2014dqa,Cheung:2015ota,Cheung:2016drk,Luo:2015tat,Kampf:2019mcd,Kampf:2023elx,Brown:2023srz,Bijnens:2019eze,Kampf:2021jvf,Low:2019ynd,Low:2022iim,Green:2022slj,Cheung:2016prv,Cheung:2022vnd,Zhou:2023quv}. NLSM amplitudes also prominently appear in other approaches -- they satisfy BCJ relations \cite{Bern:2019prr} and can be calculated using CHY formalism \cite{Cachazo:2014xea} or ambitwistor strings \cite{Casali:2015vta}, and satisfy various uniqueness conditions \cite{Arkani-Hamed:2016rak,Rodina:2016jyz,Carrasco:2019qwr}. Even more recently, exciting progress has been made on obtaining NLSM amplitudes from the surfacehedron picture \cite{Arkani-Hamed:2023lbd,Arkani-Hamed:2023mvg,Arkani-Hamed:2023swr,nima, nima2}.

In parallel, some work has gone into extending these efforts to \textit{loop integrands}. These are rational functions of the external kinematics $\lbrace p_i\rbrace$ and off-shell loop momenta $\lbrace \ell_i\rbrace$. They can be obtained as sums of Feynman diagrams prior to integration or using generalized unitarity. Unlike on-shell amplitudes, integrands are a priori not unique objects. In order to define global loop variables and meaningfully talk about a single rational function, we need to work in the planar limit $N\to \infty$ where we can perform the same decomposition as in (\ref{decomp}). This allows us to define a \emph{planar integrand} ${ I}_n^{L{\rm -loop}}$ as the kinematic coefficient of the single-trace structure.

The planar integrand is not invariant under field redefinitions. Indeed, when calculating from Feynman diagrams using flavor-ordered Feynman rules \cite{Kampf:2013vha} we get different integrands for different parametrizations. However, upon integration over loop momenta, all these integrands yield the same amplitude,
\begin{equation}
A_n^{L-{\rm loop}} = \int d^4\ell_1 d^4\ell_2 \dots d^4\ell_L\,I_n^{L-{\rm loop}}\,.
\end{equation}
The ambiguity of the loop integrand ${I}_n^{L{\rm -loop}}$ is not new; the same situation arises in planar ${\cal N}=4$ super Yang-Mills. There a \emph{preferred} integrand could be found \cite{Arkani-Hamed:2010zjl,Arkani-Hamed:2010pyv,Bourjaily:2013mma,Bourjaily:2015jna} which is invariant under a hidden \emph{dual conformal symmetry} \cite{Drummond:2007au,Drummond:2008vq}. This symmetry is also built into its construction using on-shell diagrams and the positive Grassmannians \cite{Arkani-Hamed:2012zlh} which provide triangulations of an underlying Amplituhedron geometry \cite{Arkani-Hamed:2013jha,Arkani-Hamed:2017vfh,Damgaard:2019ztj,Ferro:2022abq,Even-Zohar:2021sec,Even-Zohar:2023del,Dian:2022tpf,Arkani-Hamed:2021iya,Brown:2023mqi}. By analogy, a natural question to ask is if there exists a consistent \emph{soft theorem} for NLSM integrands that would allow to distinguish a preferred object that satisfies it.

In this letter, we provide a candidate for one such soft theorem, valid to all loops and all multiplicities, which we call the \textit{algebraic soft theorem}. Correspondingly we identify a tantalizing new recursive decomposition of NLSM tree-level off-shell correlators that manifests the new soft theorem, which itself is a natural generalization of the Adler zero to off-shell momenta. We further provide evidence that this construction can be extended to the loop level, giving a number of explicit examples. Following the statement of the algebraic soft theorem for all loops and multiplicities, we identify one solution: the (off-shell) NLSM surface integrand obtained from a certain shift of the ${\rm Tr}\phi^3$ surfacehedron \cite{Arkani-Hamed:2023lbd,Arkani-Hamed:2023mvg,Arkani-Hamed:2023swr,nima,nima2}. To conclude we derive an on-shell version of the algebraic soft theorem, revealing a rich soft structure of NLSM on-shell integrands.

\vspace{-0.4cm}

\section{Failure of Adler zero for the loop integrand}

\vspace{-0.2cm}

We start with some misleading evidence at one loop. The structure of one-loop amplitudes is sufficiently simple that we can analyze the whole space of possible loop integrands. The general decomposition gives us the sum of boxes, triangles, bubbles, tadpoles and polynomials in loop momentum. The set of polynomials also includes functions of external kinematics only. All integrands are fixed from cuts via generalized unitarity in terms of tree-level amplitudes (cut constructible part, CC), except for tadpoles and polynomials in loop momenta which do not have any physical cut, 
%
\begin{align*}
    I_n^{\rm 1-loop} = \left(\text{CC}\right) + \left( \text{tadp.} \right) + \left(\text{poly. in } \ell_k \right).
\end{align*}
%
%
Unlike in the planar ${\cal N}=4$ SYM theory where the single cut was equal to the forward limit of the tree-level amplitude \cite{Caron-Huot:2010fvq,Arkani-Hamed:2010zjl}, an analogous statement is not true in the NLSM as the forward limit of trees diverges. Furthermore, massless tadpoles and polynomials integrate to zero and can be freely added to the integrand.

In \cite{Bartsch:2022pyi}, we gave a prescription to fix tadpoles and polynomials by demanding that the planar integrand \emph{vanishes in the soft limit}, i.e. we upgraded the Adler zero to an integrand-level statement. This did not fix the integrand uniquely -- we found a one-parametric solution and imposed further constraints to obtain a unique object, for which we also formulated recursion relations. When extending this analysis to two loops, we realize that there exists \emph{no integrand} which satisfies all cuts and exhibits the Adler zero. This is already true at four points. It leads us to the conclusion that the Adler zero is not a property of planar NLSM integrands beyond one loop.

\vspace{-0.4cm}

\section{A hidden soft theorem at tree level}

\vspace{-0.3cm}

We return back to tree-level amplitudes where we want to analyze the structure of the Adler zero in more detail. The two lowest-order NLSM amplitudes are
\begin{align}
\begin{split}
    A_4 &= X_{13} + X_{24},\\
    A_6 &= \left( -\frac{1}{2} \frac{(X_{13}+X_{24})(X_{46}+X_{15})}{X_{14}} + X_{13} \right) + \text{cyc.}\, ,
    \label{NLSMboring}
\end{split}
\end{align}
where we introduced \emph{planar variables} \cite{Arkani-Hamed:2017mur}
\begin{equation}
    X_{ij} \equiv (p_i + p_{i{+}1} + \dots + p_{j{-}1})^2 \label{Xdef}.
\end{equation}
Note that $X_{i\,i{+}1} = p_i^2$ which is zero in the massless case while $X_{ii}=0$ identically. For our discussion it will be important to highlight a particular fact about the soft behavior of NLSM amplitudes that was first appreciated from the CHY perspective \cite{Cachazo:2016njl}. It was found that the coefficient of the leading vanishing term in the Adler zero is controlled by amplitudes in an extended theory coupling bi-adjoint scalars $\phi$ to NLSM pions $\pi$. Some of the simplest mixed amplitudes in this theory are
\begin{align}
M_3(1^\phi,2^\phi,3^\phi) &= 1,\label{mixedAmps}\\
M_5(1^\phi,2^\pi,3^\pi,4^\phi,5^\phi) &= 1 - \frac{X_{13}+X_{24}}{X_{14}} + \frac{X_{24}+X_{35}}{X_{25}} \nonumber.
\end{align}
Our ultimate goal is to formulate a consistent soft theorem for NLSM loop integrands. The first step we take in this direction is to promote amplitudes $A_n, M_n$ to off-shell correlators $\boldsymbol{A}_n, \boldsymbol{M}_n$. To do this, we implement a \textit{minimal prescription}. We simply declare that the algebraic form of the off-shell correlators in terms of $X_{ij}$ be the same as that of the on-shell amplitudes, i.e. we define
\begin{align}
 \boldsymbol{A}_n \equiv A_n, \hspace{0.5cm} \boldsymbol{M}_n \equiv M_n,
 \label{minPrCorr}
\end{align}
as functions of the planar variables. Notably, the correlators could in principle explicitly depend on variables $X_{ii+1}=p_i^2\neq 0$ but in the minimal prescription they do not. Thus $\boldsymbol{A}_n, \boldsymbol{M}_n$ can be interpreted as amputated Green's functions and the on-shell amplitudes can be obtained by a simple on-shell limit 
\begin{align}
    \boldsymbol{A}_n \xrightarrow[]{X_{ii{+}1}=0} A_n, \hspace{0.3cm} \boldsymbol{M}_{n} \xrightarrow[]{X_{ii{+}1}=0} M_{n}.
    \label{OSlim}
\end{align}
From the on-shell perspective, this might look like taking a step in the wrong direction. We traded unique on-shell amplitudes for some a priori ambiguous correlators. Indeed, these objects are generally not invariant under field redefinitions. Different parametrizations of the Lagrangian would produce different functions, all of which would agree only once the on-shell limit is taken. However, in this sense correlators are already more analogous to loop integrands which are the objects we are eventually interested in.

We now come to one of the central observations of this letter. To any number of points, the minimal prescription correlators (\ref{minPrCorr}) can be recursively decomposed as~\cite{noterecursion}
\begin{equation}
    \boldsymbol{A}_n = X_{1n{-}1}\,\boldsymbol{M}_{n-1}^a + X_{2n}\,\boldsymbol{M}_{n-1}^b + \boldsymbol{R}_n.
    \label{TreeBGDecomp}
\end{equation}
This representation breaks manifest cyclicity by making the $n$-th leg special. Indeed, we will show that this way of writing the correlator manifests an off-shell generalization of the Adler zero for $p_n\to 0$ and naturally implies the latter once the on-shell limit is taken. The correlators $\boldsymbol{M}_{n-1}^{a,b}$ appearing in (\ref{TreeBGDecomp}) always involve three adjacent bi-adjoint scalars (dashed lines), whereas the remaining legs are associated with pions (solid lines),
%
%
\begin{align}
\hspace{-0.2cm}\boldsymbol{M}_{n-1}^{a}= \begin{matrix} \vspace{-0.15cm} \includegraphics[width=2.0cm]{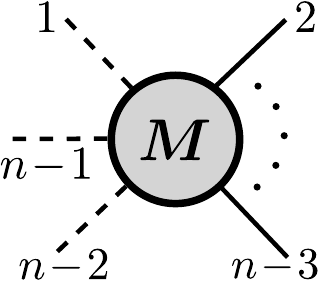}\end{matrix}\,\, , \hspace{0.5cm}  \boldsymbol{M}_{n-1}^{b}=\begin{matrix} \vspace{-0.15cm} \includegraphics[width=2.0cm]{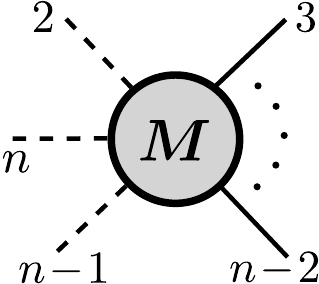} \end{matrix}\,\,.    
\end{align}
%
%
The external lines suppressed by $\dots$ correspond to pions and the labeling indicates that the middle scalar has off-shellness $(p_{n{-}1}\!+\!p_n)^2=X_{1n{-}1}$ for $\boldsymbol{M}^a$ and $(p_n\!+\!p_1)^2 = X_{2n}$ for $\boldsymbol{M}^b$ respectively. Thus $\boldsymbol{M}^a$ and $\boldsymbol{M}^b$ are related simply by a cyclic shift of the momentum labels. The \textit{remainder} $\boldsymbol{R}_n$ in (\ref{TreeBGDecomp}) can itself be decomposed further in terms of lower-point correlators in the extended theory,
\begin{align}
\begin{split}
    \boldsymbol{R}_n &= \sum_{j=2}^{n/2-1} \!\!\boldsymbol{S}(X_{1,2j},X_{2j,n})\, \boldsymbol{A}_{2j}^c\, \boldsymbol{M}_{n-2j+1}^c \\
    &+ \sum_{j=1}^{n/2-2} \!\!\boldsymbol{S}(X_{2j+1,n},X_{1,2j+1})\, \boldsymbol{M}^d_{2j+1}\, \boldsymbol{A}_{n-2j}^d.
    \label{BGtreeNLSM}
\end{split}
\end{align}
Here we have introduced the \textit{soft factor},
\begin{align}
 \boldsymbol{S}(X_A,X_B) = \frac{X_A-X_B}{X_A},
\end{align}
with the crucial property that $\boldsymbol{S}=0$ for $X_A=X_B \neq 0$. The distribution of external $\phi$ and $\pi$ particles for the correlators in (\ref{BGtreeNLSM}) is
\begin{align}
\begin{split}
\hspace{-0.2cm}\boldsymbol{A}_{2j}^c = \begin{matrix} \vspace{-0.1cm} \includegraphics[width=2.0cm]{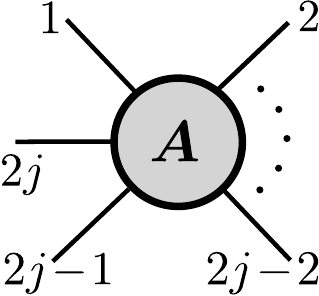}\end{matrix}, \hspace{0.2cm} \boldsymbol{A}_{n-2j}^d = \begin{matrix} \includegraphics[width=2.0cm]{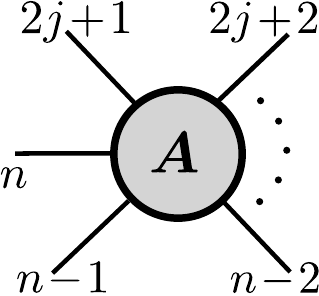}\end{matrix},\\
\hspace{-0.2cm}\boldsymbol{M}_{n-2j+1}^c = \begin{matrix} \includegraphics[width=2.0cm]{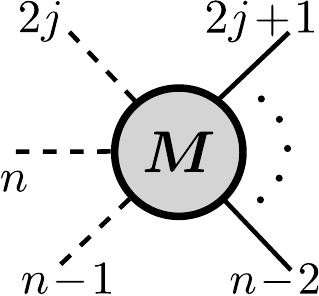}\end{matrix}, \hspace{0.2cm} \boldsymbol{M}_{2j+1}^d = \begin{matrix} \vspace{-0.1cm}\includegraphics[width=2.0cm]{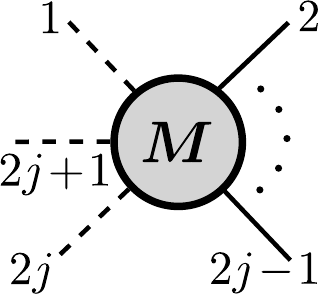}\end{matrix},
\end{split}
\end{align}
%
%
%
where again only mixed correlators with three adjacent bi-adjoint scalars appear.
Note that for general off-shell momenta
\begin{align}
   \hspace{-0.22cm} X_{in}-X_{i1}=p_n\cdot (p_1+\ldots +p_{i-1}-(p_i+\ldots +p_{n-1})).
   \label{Xsft}
\end{align}
Thus, in terms of the variables $X_{ij}$, the limit $p_n\to 0$ can be reformulated by defining an \emph{algebraic soft limit},
\begin{center}
\boxed{\mbox{Algebraic soft limit: replace label $n$ by label $1$ in $X_{ij}$.}}
\end{center}
%
Since the soft factors in the remainder (\ref{BGtreeNLSM}) are of the form (\ref{Xsft}), they ensure that it vanishes in the algebraic soft limit,
\begin{align}
    \boldsymbol{R}_n \xmapsto{n\mapsto 1} 0\,.
    \label{RalgAdler}
\end{align}
With this in mind, we can also take the algebraic soft limit of the NLSM correlator (\ref{TreeBGDecomp}) to obtain
\begin{align}
    \boldsymbol{A}_n \xmapsto{n\mapsto 1} X_{1n-1}\boldsymbol{M}^a_{n-1} + X_{12}\boldsymbol{M}^b_{n-1},
    \label{treeSftThm}
\end{align}
where $\boldsymbol{M}^{a,b}$ are now evaluated on soft kinematics $p_n\!=\!0$. For on-shell kinematics where $X_{1n-1}\!=\!X_{12}\!=\!0$ the above establishes the Adler zero of NLSM amplitudes. For off-shell kinematics $X_{1n-1}$, $X_{12}$ are non-zero and (\ref{treeSftThm}) quantifies exactly how the Adler zero fails in this case. More specifically it tells us that the leading coefficient of the (now non-vanishing) soft theorem for the off-shell minimal prescription NLSM correlators $\boldsymbol{A}_n$ is once again controlled by the mixed correlators $\boldsymbol{M}^{a,b}$. 
The uplift of the Adler zero to what we will call the \textit{algebraic soft theorem} (\ref{treeSftThm}) gives us a criterion to select a preferred form of the correlators $\boldsymbol{A}_n$ and $\boldsymbol{M}_n$, ones which satisfy (\ref{TreeBGDecomp}) and (\ref{BGtreeNLSM}). The minimal prescription correlators  $\boldsymbol{A}_n$ agree with the results obtained from the minimal parametrization Lagrangian \cite{Kampf:2013vha} (explaining the original designation). They are also identical to the objects obtained from the shifted ${\rm Tr} \phi^3$ surfacehedron amplitudes \cite{nima2}.

With an eye on later extensions to loop-level let us propose a useful interpretation of the algebraic soft theorem (\ref{treeSftThm}). To this end, we define the NLSM \textit{two-point} correlator
\begin{align}
    \begin{matrix} \vspace{-0.15cm}\includegraphics[width=1.2cm]{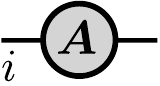}\end{matrix} = \boldsymbol{A}_2(i,i+1) \equiv X_{ii+1} = p_i^2.
    \label{defA2}
\end{align}
Notably, in the on-shell limit we have $A_2(i,i\!+\!1)$=0. With this we can identify the prefactors of $\boldsymbol{M}^{a,b}$ in (\ref{treeSftThm}),
\begin{align*}
    X_{1n-1} = \boldsymbol{A}_2(n-1,1), \hspace{0.3cm} X_{12}=\boldsymbol{A}_2(1,2).
\end{align*}
Viewed in this way, the soft theorem (\ref{treeSftThm}) can be graphically represented concisely as
%
%
%
\begin{align}
    \hspace{-0.35cm}\boldsymbol{A}_n \xmapsto{n\mapsto 1} \!\!\!\sum_{i=n-1,1}\, \begin{matrix} \vspace{-0.14cm} \includegraphics[width=3.1cm]{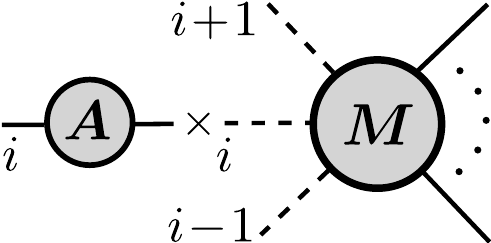}\end{matrix}\,\,,
    \label{treeSftThmDiag}
\end{align}
%
%
%
%
where, since we are on soft kinematics $p_n\!=\! 0$, the index $i$ is cyclic mod $n\!-\!1$.

Thus the NLSM correlator factorizes in the soft limit into a \textit{self-energy} (two-point function) and an extended theory correlator. There are two such contributions from the legs $p_{i}$, $(i=n\!-\!1,1)$ adjacent to the momentum $p_n$ that was taken soft. We will see that both these properties of the soft theorem are characteristic of NLSM correlators in general, also at the loop level.


To conclude, we give a few very simple examples. The four-point NLSM correlator reads
\begin{align}
    \boldsymbol{A}_4 = X_{13}\boldsymbol{M}_3^a + X_{24}\boldsymbol{M}_3^b,
    \label{A4OFF}
\end{align}
where $\boldsymbol{M}_3^{a,b} \!=\! 1$, and comparing to the general form (\ref{TreeBGDecomp}), we see that trivially $\boldsymbol{R}_4 = 0$ at four points. At six points we have
\begin{align}
    \boldsymbol{A}_6 = X_{15}\boldsymbol{M}_5^a + X_{26}\boldsymbol{M}_5^b + \boldsymbol{R}_6,
    \label{A6OFF}
\end{align}
with $\boldsymbol{M}_5^{a,b}$ formally as in (\ref{mixedAmps}), and remainder
\begin{align*}
    \boldsymbol{R}_6 = \frac{X_{14}-X_{46}}{X_{14}} (X_{13}+X_{24}) + \frac{X_{36}-X_{13}}{X_{36}}(X_{35}+X_{46}).
\end{align*}
Both (\ref{A4OFF}) and (\ref{A6OFF}) can be immediately seen to satisfy the algebraic soft theorem (\ref{treeSftThm}). Taking the on-shell limit (\ref{OSlim})
%
%
they reproduce the known NLSM amplitudes (\ref{NLSMboring}).

\vspace{-0.4cm}

\section{The algebraic soft theorem for loop integrands}

\vspace{-0.2cm}

We now wish to define a preferred off-shell integrand, or amputated loop-level correlator ${\bf I}_n$, from which an on-shell integrand $I_n$ can be derived by taking the appropriate on-shell limit. Following the tree-level philosophy for the correlator $\boldsymbol{A}_n$ we want to look for an off-shell soft theorem, possibly a generalization of (\ref{treeSftThm}), that would be satisfied by ${\bf I}_n$. At the tree level, a crucial ingredient to ensuring the validity of the algebraic soft theorem was the rigid recursive structure of the correlator following from (\ref{TreeBGDecomp}) and (\ref{BGtreeNLSM}). Indeed, a generalization of this structure remains valid to all loop orders \cite{inprogress} provided we add one more layer of generalization: we need to allow our integrand to depend on the planar variables $X_{ii}\neq 0$. Following the definition (\ref{Xdef}) these parameters are identically zero even for off-shell kinematics, so we have to consider them as formal deformations. We refer to this integrand with explicit dependence on $X_{ii}$ as a \textit{surface integrand} and denote it by $\boldsymbol{\mathcal{I}}_n$. We can then define an on-shell integrand $I_n$ as a sequence of limits,
\begin{equation}
\boldsymbol{\mathcal{I}}_n \xrightarrow[]{X_{ii}=0} {\bf I}_n \xrightarrow[]{X_{ii{+}1}=0} I_n.
\end{equation}
As a minor caveat we should mention that the first step in this sequence, i.e. going from $\boldsymbol{\mathcal{I}}_n$ to $\boldsymbol{I}_n$, is more subtle than indicated as it will involve \textit{amputation} of external legs. We will touch upon the details of this procedure shortly. We are now ready for the main statement of this section. To any number of loops and multiplicities there exist surface functions $\boldsymbol{\mathcal{U}}_{n-1}^{a,b}$ such that the NLSM surface integrand $\boldsymbol{\mathcal{I}}_{n}$  satisfies the \textit{algebraic soft theorem}
\begin{align}
    \boldsymbol{\mathcal{I}}_n \xmapsto{n\mapsto 1} X_{1n-1}\,\boldsymbol{\mathcal{U}}^{a}_{n-1} + X_{12}\,\boldsymbol{\mathcal{U}}^{b}_{n-1}.
    \label{sftThmLoop}
\end{align}
While a fully satisfactory interpretation of the odd-point functions $\boldsymbol{\mathcal{U}}^{a,b}$ is currently lacking, their analytic form suggests (see example) that they are appropriate generalizations of the mixed theory correlators $\boldsymbol{M}^{a,b}$ to loop-level. We leave a detailed investigation for future work. When searching for the integrand that satisfies (\ref{sftThmLoop}) we find a solution: the NLSM surface integrand obtained from a shift of the ${\rm Tr} \phi^3$ surfacehedron integrand \cite{nima,nima2}. By `surface integrand' we will broadly refer to any integrand with $X_{ii}\neq 0$ (as the motivation to turn these variables on comes from the surfacehedron picture). We do not know if the shifted surfacehedron integrand is the unique solution to the soft theorem constraint (\ref{sftThmLoop}) or if more solutions exist, but some other natural candidates do fail that condition. For example, the surface integrand obtained from the minimal parametrization does not satisfy (\ref{sftThmLoop}) as it does not make use of the $X_{ii}$ variables. 

It is instructive to show why turning on $X_{ii}$ is required to satisfy the soft theorem (\ref{sftThmLoop}). Let us consider the following soft factor
\begin{align*}
\boldsymbol{S}(X_{1n},X_{11})=\frac{X_{1n}-X_{11}}{X_{1n}}.
\end{align*}
For the on-shell integrand $I_n$ this quantity is not even well-defined as $X_{1n}=0$. For the off-shell integrand $\boldsymbol{I}_n$ we have $X_{11}=0$ and the factor becomes $X_{1n}/X_{1n}=1$. Taking the algebraic soft limit, the constant factor remains unchanged. Finally, for the surface integrand $\boldsymbol{\mathcal{I}}_n$, taking the algebraic soft limit $n\mapsto 1$, we map $X_{1n} \mapsto X_{11}$ in the numerator and the soft factor vanishes as it did in the tree-level case. Exactly these types of cancellations are needed for the surface integrand $\boldsymbol{\mathcal{I}}_n$ to satisfy (\ref{sftThmLoop}).
Let us now illustrate our assertions on simple examples.

\section{Two- and four-point examples}

\vspace{-0.2cm}

Here we present results for the simplest two- and four-point one-loop surface integrands. At the loop level, integrands depend on the planar loop variables,
\begin{align}
    X_{iz} = \left( \ell + p_1 + \ldots + p_{i-1} \right)^2,
\end{align}
where we denote by $\ell$ and $z$ the loop momentum and its associated label accordingly. We start with the two-point surface integrand, or \textit{self-energy} which we can decompose in a way analogous to the tree-level correlator 
(\ref{TreeBGDecomp}),
\begin{align}
\begin{split}
    \boldsymbol{\mathcal{I}}_2(1,2) &= X_{11}\,\boldsymbol{\mathcal{U}}_1\!(1) + X_{22}\,\boldsymbol{\mathcal{U}}_1\!(2) + \boldsymbol{\mathcal{R}}_2\label{SEandTAD},\\
\end{split}
\end{align}
in terms of the one-loop \textit{tadpole},
\begin{align}
\boldsymbol{\mathcal{U}}_1(1) = -\frac{1}{X_{1z}},
\end{align}
and a remainder
\begin{align}
     \boldsymbol{\mathcal{R}}_2 &= \boldsymbol{S}(X_{2z},X_{1z}) + \boldsymbol{S}(X_{1z},X_{2z}).
\end{align}
Since $\boldsymbol{\mathcal{R}}_2$, like its tree-level analogue (\ref{BGtreeNLSM}), is composed of soft factors, it vanishes in the algebraic soft limit, that is, $\boldsymbol{\mathcal{R}}_2 \mapsto 0$ for $2 \mapsto 1$.
%
%
This ensures that $\boldsymbol{\mathcal{I}}_2$ satisfies the algebraic soft theorem (\ref{sftThmLoop}). Moving on to four points we observe that the integrand still follows the expected structure,
\begin{align}
   \boldsymbol{\mathcal{I}}_4 = X_{13}\,\boldsymbol{\mathcal{U}}_3^a +  X_{24}\,\boldsymbol{\mathcal{U}}_3^b  + \boldsymbol{\mathcal{R}}_4,
   \label{I41L}
\end{align}
where now
\begin{align*}
\begin{split}
    \boldsymbol{\mathcal{U}}_3^a = \boldsymbol{\mathcal{U}}_3\!(1,2,3) &= \frac{X_{13}+X_{2z}}{X_{1z}X_{3z}} - \frac{1}{X_{1z}} - \frac{1}{X_{2z}} - \frac{1}{X_{3z}}\\ 
    &- \frac{1}{X_{12}} \boldsymbol{\mathcal{I}}_2(1,2) - \frac{1}{X_{23}} \boldsymbol{\mathcal{I}}_2(2,3),
\end{split}
\end{align*}
and $\boldsymbol{\mathcal{U}}_3^{b}=\boldsymbol{\mathcal{U}}_3(2,3,1)$. 
%
%
%
The remainder $\boldsymbol{\mathcal{R}}_{4}$ is given by
%
%
\begin{align*}
    \boldsymbol{\mathcal{R}}_4 &= \boldsymbol{S}(X_{4z},X_{1z})\!\left(\! 1 - \frac{X_{13}+X_{24}}{X_{14}} - \frac{X_{24}+X_{3z}}{X_{2z}}  \!\right) \label{R41}\\
    &+\boldsymbol{S}(X_{12},X_{24})\,\boldsymbol{\mathcal{M}}_3\,\boldsymbol{\mathcal{I}}_2(1,2) + \boldsymbol{S}(X_{14},X_{11})\,\boldsymbol{\mathcal{A}}_4\,\boldsymbol{\mathcal{U}}_1\!(1) \nonumber\\
    &+ (1\leftrightarrow 4,\, 2\leftrightarrow 3),\nonumber
\end{align*}
with surface three- and four-point functions $\boldsymbol{\mathcal{M}}_3 =1$, $\boldsymbol{\mathcal{A}}_4 = X_{13} + X_{24}$ identical to their correlator counterparts. The notation $(i\leftrightarrow j, \dots)$ indicates to add the same set of terms with momentum labels permuted accordingly. 

Again, the soft factor structure of $\boldsymbol{\mathcal{R}}_4$ manifests the fact that $\boldsymbol{\mathcal{R}}_4 \mapsto 0$ in the algebraic soft limit $4 \mapsto 1$,
%
%
which in turn implies the validity of the algebraic soft theorem (\ref{sftThmLoop}) for $\boldsymbol{\mathcal{I}}_{4}$.
%
%
The above examples of $\boldsymbol{\mathcal{I}}_2$ and $\boldsymbol{\mathcal{I}}_4$ clearly reveal a recursive structure of the surface integrands, closely following the tree-level pattern of (\ref{TreeBGDecomp}) and (\ref{BGtreeNLSM}), which will be the subject of future work \cite{inprogress}.

To wrap up this section, two remarks about the general structure of the surface correlators $\boldsymbol{\mathcal{I}}_{n}$ are in order. Firstly, although not at all manifest, the correlators are cyclic in the external labels $\lbrace 1,\ldots,n \rbrace$. Secondly, the correlators include contributions from self-energy corrections on external legs. E.g. in our four-point example, the integrand contains a structure
\begin{align}
    \boldsymbol{\mathcal{I}}_{4} \subset \boldsymbol{\mathcal{I}}_{2}\frac{1}{X_{12}}\boldsymbol{\mathcal{A}}_{4}= \begin{matrix} \vspace{-0.1cm} \includegraphics[width=3.0cm]{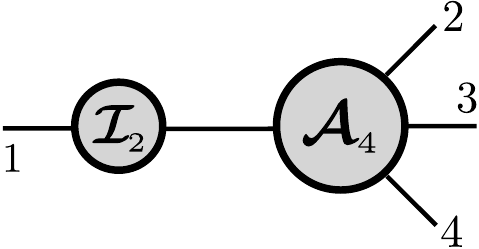} \end{matrix}\,\,.
\end{align}
The surface integrands $\boldsymbol{\mathcal{I}}_{n}$ therefore correspond to certain deformations (recall $X_{ii}\neq 0$) of \textit{partially amputated} (i.e. without free propagators on external legs) NLSM Green's functions. A fact that will be of prime importance as we want to transition to the on-shell limit next.

\vspace{-0.3cm}

\section{Back to on-shell integrands}

\vspace{-0.2cm}

We would now like to extract an on-shell integrand $I_n$ from the surface object $\boldsymbol{\mathcal{I}}_n$. After all, $I_n$ is the function that needs to be integrated to obtain the corresponding loop amplitude. As we have seen, the surface integrand $\boldsymbol{\mathcal{I}}_n$ includes self-energy corrections on external legs. This was necessary to ensure the validity of the algebraic soft theorem (\ref{sftThmLoop}). Of course, these terms are divergent in the on-shell limit as they contain (powers of) propagators $X_{ii+1}^{-1}$. We will \textit{amputate} these contributions, effectively just dropping them from $\boldsymbol{\mathcal{I}}_n$, to make the on-shell limit well-defined. In the same step we turn off the deformation parameters $X_{ii}\equiv 0$ to obtain the amputated off-shell correlator $\boldsymbol{I}_n$.
%
%
Finally, taking the (now smooth) on-shell limit $X_{ii+1}\equiv 0$ we arrive at the on-shell integrand $I_n$. 

Let us again start with the simplest example of the one-loop two-point function (\ref{SEandTAD}). In this case there is nothing to amputate and we can immediately take $X_{ii},X_{ii+1}\to 0$. Thus we obtain the on-shell self-energy
\begin{align}
    I_2(1,2) &= \boldsymbol{S}(X_{2z},X_{1z}) + \boldsymbol{S}(X_{1z},X_{2z}).
    \label{UIUdefON}
\end{align}

At four-points, the one-loop on-shell integrand $I_4$ extracted from the surface function $\boldsymbol{\mathcal{I}}_{4}$ in (\ref{I41L}) reads
\begin{align}
I_4 &= X_{13}U_3^a + X_{24}U_3^b + R_4   \label{I41LON}\\
&+  I_2(1,2)M_3 + I_2(3,4)M_3 + A_4\, U_1\!(1) + A_4\, U_1\!(4). \nonumber
\end{align}
The process of amputation and going on-shell has left us with the lower-point on-shell objects
\begin{align*}
    U_3^a &= U_3\!(1,2,3) = \frac{X_{13}+X_{2z}}{X_{1z}X_{3z}} - \frac{1}{X_{1z}} - \frac{1}{X_{2z}} - \frac{1}{X_{3z}},
\end{align*}
as well as $U_3^b = U_3(2,3,1)$ and $U_1\!(1) = \boldsymbol{\mathcal{U}}_1(1)$.
In (\ref{I41LON}) we have chosen to separate the self-energy and tadpole contributions from the on-shell remainder $R_4$. This was done since, comparing to the surface remainder (\ref{R41}), the soft factors corresponding to these terms get partially amputated in the on-shell limit and thus their soft properties are spoiled. E.g. in the on-shell limit
\begin{align}
\begin{split}
    \boldsymbol{\mathcal{R}}_4 &\supset \boldsymbol{S}(X_{12},X_{24}) \mapsto 1.
    \label{sftFactAmp}
\end{split}
\end{align}
Explicitly excluding these terms from the definition of the on-shell remainder $R_4$ leaves us with
%
\begin{align*}
\begin{split}
    R_4 &= \boldsymbol{S}(X_{4z},X_{1z})\! \left(\! 1 - \frac{X_{24}+X_{3z}}{X_{2z}} \!\right) + (1\leftrightarrow 4,\, 2\leftrightarrow 3).
\end{split}
\end{align*}
This ensures that it still satisfies the soft limit $R_4 \xrightarrow{p_4\to 0} 0$
%
in close analogy with (\ref{RalgAdler}).
However, from (\ref{I41LON}) it can be seen that the integrand as a whole does not exhibit the Adler zero, precisely as a consequence of the partial amputation (\ref{sftFactAmp}). Instead, for $p_4\rightarrow0$ we get 
\begin{align}
\lim_{p_4\to 0} I_4 = I_2(1,2) M_3  + I_2(3,1) M_3,
\label{I41sft1}
\end{align}
as all other terms in (\ref{I41LON}) manifestly vanish.
%
%
%
Just as for the tree-level correlator (\ref{treeSftThm}) the result splits into the \emph{self-energy} $I_2$ and \emph{mixed amplitudes} $M_3$. This generalizes to the one-loop $n$-point case where the on-shell soft theorem takes the familiar form
%
%
%
%
%
\begin{align}
    \lim_{p_n\to 0} I_n = \sum_{i=n-1,1} \begin{matrix} \vspace{-0.1cm} \includegraphics[width=3.0cm]{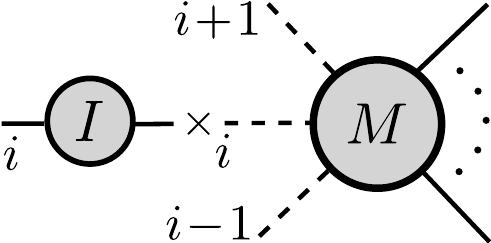}\end{matrix}\,\,,
    \label{ONsft1loop}
\end{align}
with the kinematic configuration of self-energies $I_2$ and amplitudes $M_{n-1}$ as previously in (\ref{treeSftThmDiag}). Note that in (\ref{ONsft1loop}) we are strictly in the soft limit $p_n\!=\! 0$ and as such $M_{n-1}$ correspond to true on-shell amplitudes.

\vspace{-0.3cm}

\section{All-loop on-shell soft theorem}

\vspace{-0.2cm}

At higher loops ($L\ge 2$) the full \textit{all-loop, all-point} structure of the on-shell soft theorem emerges. Schematically, the soft theorem can always be written as a sum over contributions, each factorizing into three parts,
\begin{align}
    \lim_{p_n\to 0} I_n^{(L)} \simeq \sum \,\,\,\, \begin{matrix} \vspace{-0.14cm} \includegraphics[width=3.9cm]{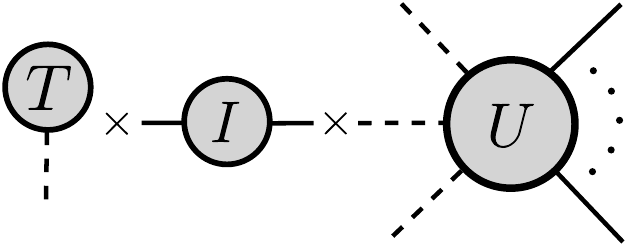} \end{matrix}.
    \label{ONsftLloop1}
\end{align}
Each term involves a \textit{tadpole factor} $T_1$, a self-energy $I_2$ and a $(n\!-\!1)$-point integrand $U_{n-1}$. The higher-loop integrands $I_2$ and $U_{n-1}$ are analogues of the on-shell objects derived from surface functions as shown previously. The tadpole factor $T_1$ can be constructed from tadpole functions $U_1$ as we will discuss shortly.

Since each factor in (\ref{ONsftLloop1}) has a given loop-order $k_j$, $j\!=\!T,I,U$, we can write the soft theorem more precisely
\begin{align}
 \lim_{p_n\to 0} I_n^{(L)} = \!\!\!\!\!\sum_{k_T+k_I+k_U=L}\!\!\!\!\! T_1^{(k_T)} \times \!\!\!\!\!\sum_{i=n-1,1} I_2^{(k_I)}\times U_{n-1}^{(k_U)},
 \label{ONsftLloop3}
\end{align}
where we sum over all triplets of integers $(k_T,k_I,k_U)\ge 0$ such that $k_T\!+\!k_I\!+\!k_U\!=\!L$. Also included is the usual sum over adjacent legs $p_i$, with $i\!=\!n\!-\!1,1$. We note that the tadpole factors $T_1$ are identical for both of these contributions. The kinematic dependence of $I_2$ and $U_{n-1}$ in (\ref{ONsftLloop3}) is exactly as in the previously established soft theorems (\ref{treeSftThmDiag}) and (\ref{ONsft1loop}) and we have $T_1\!=\!T_1(1)$.
Let us first comment on the cases where any of the $k_j=0$. We define
\begin{align}
 T_1^{(0)} \equiv 1, \hspace{0.5cm} I_2^{(0)} \equiv A_2 = 0, \hspace{0.5cm} U_{n-1}^{(0)} \equiv M_{n-1}.
 \label{TIU0def}
\end{align}
This ensures that (\ref{ONsftLloop3}) correctly reproduces the already established results for the cases $L\!=\!0,1$. Indeed, at tree-level $(L\!=\!0)$ only one triplet $(k_T,k_I,k_U)=(0,0,0)$ contributes. In particular, due to (\ref{TIU0def}), the result will be proportional to $A_2=0$. In this way we recover the Adler zero for tree amplitudes. At one loop ($L=1$) there is only one non-trivial contribution from the triplet $(k_T,k_I,k_U)=(0,1,0)$ which gives the soft theorem (\ref{ONsft1loop}).

Before we move on to higher loops, let us state the definition of the tadpole factor. For $L\ge 1$ it is given by
\begin{align}
    T_1^{(L)} = \sum_{K=1}^{L}2^K \!\!\!\!\!\!\sum\limits_{k_1+\dots+k_K=L} \!\!\!\!\!\! U_1^{(k_1)}U_1^{(k_2)}\ldots \,U_1^{(k_K)}\,,
    \label{Tdef}
\end{align}
where for fixed $K$ the second sum runs over all integers $(k_1,k_2,\ldots,k_K)\ge 0$ with $k_1\!+k_2\!+\ldots\!+\!k_K\!=\!L$. Similarly to the sum in the soft theorem (\ref{ONsftLloop3}), here we sum over all possible ways to distribute the $L$ loops among the $K$ factors of $U_1^{(k_j)}$ in (\ref{Tdef}). For the cases where any of the $k_j\!=\!0$ we define $U_1^{(0)}\equiv 0$.

To give an explicit example, at two loops the soft theorem (\ref{ONsftLloop3}) has three non-trivial contributions from $(k_T,k_I,k_U)=(0,2,0),(1,1,0),$ and $(0,1,1)$. Graphically they can be represented as
\begin{align*}
\begin{split}
    \lim_{p_n\to 0} I_n^{(2)} \simeq\,\, \begin{matrix} \vspace{-0.14cm} \includegraphics[width=5.8cm]{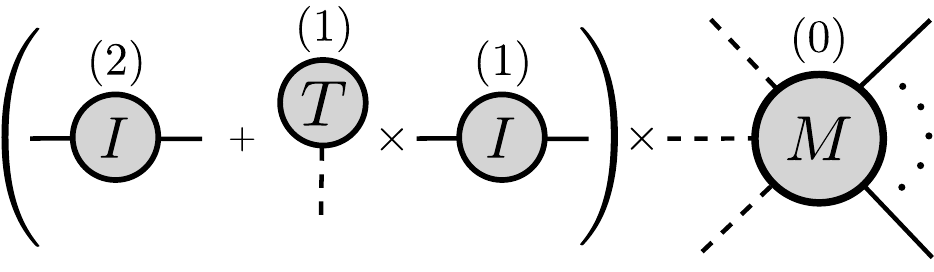} \end{matrix}\\
    + \,\, \begin{matrix} \vspace{-0.05cm} \includegraphics[width=3.0cm]{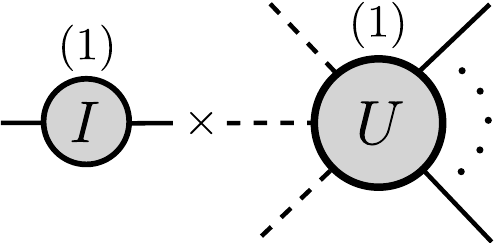} \end{matrix}
\end{split}
\end{align*}
where the brackets over the diagrams indicate loop orders and we have combined contributions by fixed loop order $k_U=0,1$. The sum over $i\!=\!n\!-\!1,1$ is implicit. 

Finally, let us mention one curious exception to the general soft theorem (\ref{ONsftLloop3}). For the case $n=2$ the Adler zero is in fact preserved to all loop orders and we have
\begin{align*}
    \lim_{p_1\to 0} I_2^{(L)} = 0.
\end{align*}
We leave for future work \cite{inprogress} how this follows from the surface integrand construction \cite{nima,nima2}. However, it can be readily verified at $L=1$ for our example (\ref{UIUdefON}).

\vspace{-0.4cm}

\section{Conclusion and Outlook}

\vspace{-0.2cm}
In this letter we proposed a consistent soft theorem for planar NLSM loop integrands with any number of loops and legs, called the \textit{algebraic soft theorem}. While tree-level amplitudes enjoy the Adler zero, the property is generically violated for loop integrands and only restored after integration. To find a simple non-vanishing soft theorem for integrands we generalized to off-shell correlators and considered a certain deformation of kinematics. We identified a new recursive structure of tree-level correlators and defined a simple kinematical operation, an \textit{algebraic soft limit}, which reduces to the usual soft limit for on-shell momenta. In this setup, both tree-level and deformed loop (surface) correlators satisfy the surprisingly simple algebraic soft theorem. They also agree with the functions obtained from the surfacehedron picture \cite{nima, nima2}. We then derived the on-shell version of the algebraic soft theorem, which is non-zero (due to the violation of the Adler zero by \textit{amputation}) but nicely factorizes into three parts: tadpole functions, self-energies and lower-point integrands in the mixed scalar-pion theory. Our result poses the question if the Adler zero can be restored for on-shell integrands using some scheme which would account for the external leg renormalization. 

NLSM amplitudes belong to a larger family of mixed scalar-pion amplitudes which are hidden in their soft structure. This connection was first established for tree-level amplitudes using the CHY formalism \cite{Cachazo:2014xea}. Similarly, the algebraic soft theorem is governed by the same mixed amplitudes and their loop-level analogues, whose precise interpretation we leave for future work \cite{inprogress}. 

Further questions involve adding masses, higher-derivative operators and a more general understanding of the surface integrands in other theories. It would also be interesting to explore integrands for theories without ordering. Natural starting points are loop amplitudes for theories with extended soft limits such as Dirac-Born-Infeld theory or the special Galileon \cite{Cheung:2014dqa,Cachazo:2014xea,Casali:2015vta,Hinterbichler:2015pqa,Cheung:2016drk,Novotny:2016jkh,Cheung:2018oki,Elvang:2018dco,Roest:2019oiw,Preucil:2019nxt,Kampf:2020tne,Brauner:2020ezm,Kampf:2021bet,deNeeling:2022tsu}.

\medskip

{\it Acknowledgements}: We thank Nima Arkani-Hamed, Taro Brown, Carolina Figueiredo, Song He, Henrik Johansson, Umut Oktem and Shruti Paranjape for very useful discussions. This work is supported by GA\-\v{C}R 24-11722S, GAUK-327422, MEYS LUAUS23126, OP JAK CZ.02.01.01/00/22\_008/0004632, DOE grant No. SC0009999 and the funds of the University of California.
\bibliography{mainbib}
\bibliographystyle{apsrev4-1}

\end{document}

%% file: header_data.tex
\titleformat{\section}{\centering\normalsize\normalfont\bf}{\thesection}{0em}{}
\hypersetup{pdftitle={},pdfcreator={},linkcolor=[rgb]{0.15,0.35,0.75},colorlinks=true,citecolor=[rgb]{0.675,0,0.2},urlcolor=[rgb]{0.15,0.35,0.65}}
\renewcommand{\phi}{\varphi}

\definecolor{nmhv}{rgb}{0.95,0.55,0.55}
\definecolor{mhvblue}{rgb}{0.6,0.6,0.7765}
\definecolor{ampgrey}{rgb}{0.9,0.9,0.9}
\definecolor{hblue}{rgb}{0,0,0.575}
\definecolor{hred}{rgb}{0.575,0.0,0.225}
\definecolor{hgreen}{rgb}{0.0,0.4,0.2}
\definecolor{hteal}{rgb}{0.0,0.545,0.7451}
\definecolor{perm}{rgb}{0.1,0.45,0.85}
\definecolor{unord}{rgb}{0,0,0}
\definecolor{ord}{rgb}{0,0,0.575}
\definecolor{anchorLeg}{rgb}{0.575,0.0,0.225}
\definecolor{fRed}{rgb}{0.48,0.02824,0.18824}

%% file: main.bbl
\begin{thebibliography}{60}%
\makeatletter
\providecommand \@ifxundefined [1]{%
 \@ifx{#1\undefined}
}%
\providecommand \@ifnum [1]{%
 \ifnum #1\expandafter \@firstoftwo
 \else \expandafter \@secondoftwo
 \fi
}%
\providecommand \@ifx [1]{%
 \ifx #1\expandafter \@firstoftwo
 \else \expandafter \@secondoftwo
 \fi
}%
\providecommand \natexlab [1]{#1}%
\providecommand \enquote  [1]{``#1''}%
\providecommand \bibnamefont  [1]{#1}%
\providecommand \bibfnamefont [1]{#1}%
\providecommand \citenamefont [1]{#1}%
\providecommand \href@noop [0]{\@secondoftwo}%
\providecommand \href [0]{\begingroup \@sanitize@url \@href}%
\providecommand \@href[1]{\@@startlink{#1}\@@href}%
\providecommand \@@href[1]{\endgroup#1\@@endlink}%
\providecommand \@sanitize@url [0]{\catcode `\\12\catcode `\$12\catcode `\&12\catcode `\#12\catcode `\^12\catcode `\_12\catcode `\%12\relax}%
\providecommand \@@startlink[1]{}%
\providecommand \@@endlink[0]{}%
\providecommand \url  [0]{\begingroup\@sanitize@url \@url }%
\providecommand \@url [1]{\endgroup\@href {#1}{\urlprefix }}%
\providecommand \urlprefix  [0]{URL }%
\providecommand \Eprint [0]{\href }%
\providecommand \doibase [0]{http://dx.doi.org/}%
\providecommand \selectlanguage [0]{\@gobble}%
\providecommand \bibinfo  [0]{\@secondoftwo}%
\providecommand \bibfield  [0]{\@secondoftwo}%
\providecommand \translation [1]{[#1]}%
\providecommand \BibitemOpen [0]{}%
\providecommand \bibitemStop [0]{}%
\providecommand \bibitemNoStop [0]{.\EOS\space}%
\providecommand \EOS [0]{\spacefactor3000\relax}%
\providecommand \BibitemShut  [1]{\csname bibitem#1\endcsname}%
\let\auto@bib@innerbib\@empty
\bibitem [{\citenamefont {Kampf}\ \emph {et~al.}(2013{\natexlab{a}})\citenamefont {Kampf}, \citenamefont {Novotny},\ and\ \citenamefont {Trnka}}]{Kampf:2012fn}%
  \BibitemOpen
  \bibfield  {author} {\bibinfo {author} {\bibfnamefont {K.}~\bibnamefont {Kampf}}, \bibinfo {author} {\bibfnamefont {J.}~\bibnamefont {Novotny}}, \ and\ \bibinfo {author} {\bibfnamefont {J.}~\bibnamefont {Trnka}},\ }\href {\doibase 10.1103/PhysRevD.87.081701} {\bibfield  {journal} {\bibinfo  {journal} {Phys. Rev. D}\ }\textbf {\bibinfo {volume} {87}},\ \bibinfo {pages} {081701} (\bibinfo {year} {2013}{\natexlab{a}})},\ \Eprint {http://arxiv.org/abs/1212.5224} {arXiv:1212.5224} \BibitemShut {NoStop}%
\bibitem [{\citenamefont {Kampf}\ \emph {et~al.}(2013{\natexlab{b}})\citenamefont {Kampf}, \citenamefont {Novotny},\ and\ \citenamefont {Trnka}}]{Kampf:2013vha}%
  \BibitemOpen
  \bibfield  {author} {\bibinfo {author} {\bibfnamefont {K.}~\bibnamefont {Kampf}}, \bibinfo {author} {\bibfnamefont {J.}~\bibnamefont {Novotny}}, \ and\ \bibinfo {author} {\bibfnamefont {J.}~\bibnamefont {Trnka}},\ }\href {\doibase 10.1007/JHEP05(2013)032} {\bibfield  {journal} {\bibinfo  {journal} {JHEP}\ }\textbf {\bibinfo {volume} {05}},\ \bibinfo {pages} {032} (\bibinfo {year} {2013}{\natexlab{b}})},\ \Eprint {http://arxiv.org/abs/1304.3048} {arXiv:1304.3048} \BibitemShut {NoStop}%
\bibitem [{\citenamefont {Cheung}\ \emph {et~al.}(2015)\citenamefont {Cheung}, \citenamefont {Kampf}, \citenamefont {Novotny},\ and\ \citenamefont {Trnka}}]{Cheung:2014dqa}%
  \BibitemOpen
  \bibfield  {author} {\bibinfo {author} {\bibfnamefont {C.}~\bibnamefont {Cheung}}, \bibinfo {author} {\bibfnamefont {K.}~\bibnamefont {Kampf}}, \bibinfo {author} {\bibfnamefont {J.}~\bibnamefont {Novotny}}, \ and\ \bibinfo {author} {\bibfnamefont {J.}~\bibnamefont {Trnka}},\ }\href {\doibase 10.1103/PhysRevLett.114.221602} {\bibfield  {journal} {\bibinfo  {journal} {Phys. Rev. Lett.}\ }\textbf {\bibinfo {volume} {114}},\ \bibinfo {pages} {221602} (\bibinfo {year} {2015})},\ \Eprint {http://arxiv.org/abs/1412.4095} {arXiv:1412.4095} \BibitemShut {NoStop}%
\bibitem [{\citenamefont {Cheung}\ \emph {et~al.}(2016)\citenamefont {Cheung}, \citenamefont {Kampf}, \citenamefont {Novotny}, \citenamefont {Shen},\ and\ \citenamefont {Trnka}}]{Cheung:2015ota}%
  \BibitemOpen
  \bibfield  {author} {\bibinfo {author} {\bibfnamefont {C.}~\bibnamefont {Cheung}}, \bibinfo {author} {\bibfnamefont {K.}~\bibnamefont {Kampf}}, \bibinfo {author} {\bibfnamefont {J.}~\bibnamefont {Novotny}}, \bibinfo {author} {\bibfnamefont {C.-H.}\ \bibnamefont {Shen}}, \ and\ \bibinfo {author} {\bibfnamefont {J.}~\bibnamefont {Trnka}},\ }\href {\doibase 10.1103/PhysRevLett.116.041601} {\bibfield  {journal} {\bibinfo  {journal} {Phys. Rev. Lett.}\ }\textbf {\bibinfo {volume} {116}},\ \bibinfo {pages} {041601} (\bibinfo {year} {2016})},\ \Eprint {http://arxiv.org/abs/1509.03309} {arXiv:1509.03309} \BibitemShut {NoStop}%
\bibitem [{\citenamefont {Cheung}\ \emph {et~al.}(2017)\citenamefont {Cheung}, \citenamefont {Kampf}, \citenamefont {Novotny}, \citenamefont {Shen},\ and\ \citenamefont {Trnka}}]{Cheung:2016drk}%
  \BibitemOpen
  \bibfield  {author} {\bibinfo {author} {\bibfnamefont {C.}~\bibnamefont {Cheung}}, \bibinfo {author} {\bibfnamefont {K.}~\bibnamefont {Kampf}}, \bibinfo {author} {\bibfnamefont {J.}~\bibnamefont {Novotny}}, \bibinfo {author} {\bibfnamefont {C.-H.}\ \bibnamefont {Shen}}, \ and\ \bibinfo {author} {\bibfnamefont {J.}~\bibnamefont {Trnka}},\ }\href {\doibase 10.1007/JHEP02(2017)020} {\bibfield  {journal} {\bibinfo  {journal} {JHEP}\ }\textbf {\bibinfo {volume} {02}},\ \bibinfo {pages} {020} (\bibinfo {year} {2017})},\ \Eprint {http://arxiv.org/abs/1611.03137} {arXiv:1611.03137} \BibitemShut {NoStop}%
\bibitem [{\citenamefont {Luo}\ and\ \citenamefont {Wen}(2016)}]{Luo:2015tat}%
  \BibitemOpen
  \bibfield  {author} {\bibinfo {author} {\bibfnamefont {H.}~\bibnamefont {Luo}}\ and\ \bibinfo {author} {\bibfnamefont {C.}~\bibnamefont {Wen}},\ }\href {\doibase 10.1007/JHEP03(2016)088} {\bibfield  {journal} {\bibinfo  {journal} {JHEP}\ }\textbf {\bibinfo {volume} {03}},\ \bibinfo {pages} {088} (\bibinfo {year} {2016})},\ \Eprint {http://arxiv.org/abs/1512.06801} {arXiv:1512.06801} \BibitemShut {NoStop}%
\bibitem [{\citenamefont {Kampf}\ \emph {et~al.}(2020)\citenamefont {Kampf}, \citenamefont {Novotny}, \citenamefont {Shifman},\ and\ \citenamefont {Trnka}}]{Kampf:2019mcd}%
  \BibitemOpen
  \bibfield  {author} {\bibinfo {author} {\bibfnamefont {K.}~\bibnamefont {Kampf}}, \bibinfo {author} {\bibfnamefont {J.}~\bibnamefont {Novotny}}, \bibinfo {author} {\bibfnamefont {M.}~\bibnamefont {Shifman}}, \ and\ \bibinfo {author} {\bibfnamefont {J.}~\bibnamefont {Trnka}},\ }\href {\doibase 10.1103/PhysRevLett.124.111601} {\bibfield  {journal} {\bibinfo  {journal} {Phys. Rev. Lett.}\ }\textbf {\bibinfo {volume} {124}},\ \bibinfo {pages} {111601} (\bibinfo {year} {2020})},\ \Eprint {http://arxiv.org/abs/1910.04766} {arXiv:1910.04766} \BibitemShut {NoStop}%
\bibitem [{\citenamefont {Kampf}\ \emph {et~al.}(2023)\citenamefont {Kampf}, \citenamefont {Novotny}, \citenamefont {Trnka},\ and\ \citenamefont {Vasko}}]{Kampf:2023elx}%
  \BibitemOpen
  \bibfield  {author} {\bibinfo {author} {\bibfnamefont {K.}~\bibnamefont {Kampf}}, \bibinfo {author} {\bibfnamefont {J.}~\bibnamefont {Novotny}}, \bibinfo {author} {\bibfnamefont {J.}~\bibnamefont {Trnka}}, \ and\ \bibinfo {author} {\bibfnamefont {P.}~\bibnamefont {Vasko}},\ }\href {\doibase 10.1007/JHEP12(2023)135} {\bibfield  {journal} {\bibinfo  {journal} {JHEP}\ }\textbf {\bibinfo {volume} {12}},\ \bibinfo {pages} {135} (\bibinfo {year} {2023})},\ \Eprint {http://arxiv.org/abs/2303.14761} {arXiv:2303.14761} \BibitemShut {NoStop}%
\bibitem [{\citenamefont {Brown}\ \emph {et~al.}(2023{\natexlab{a}})\citenamefont {Brown}, \citenamefont {Kampf}, \citenamefont {Oktem}, \citenamefont {Paranjape},\ and\ \citenamefont {Trnka}}]{Brown:2023srz}%
  \BibitemOpen
  \bibfield  {author} {\bibinfo {author} {\bibfnamefont {T.~V.}\ \bibnamefont {Brown}}, \bibinfo {author} {\bibfnamefont {K.}~\bibnamefont {Kampf}}, \bibinfo {author} {\bibfnamefont {U.}~\bibnamefont {Oktem}}, \bibinfo {author} {\bibfnamefont {S.}~\bibnamefont {Paranjape}}, \ and\ \bibinfo {author} {\bibfnamefont {J.}~\bibnamefont {Trnka}},\ }\href {\doibase 10.1103/PhysRevD.108.105008} {\bibfield  {journal} {\bibinfo  {journal} {Phys. Rev. D}\ }\textbf {\bibinfo {volume} {108}},\ \bibinfo {pages} {105008} (\bibinfo {year} {2023}{\natexlab{a}})},\ \Eprint {http://arxiv.org/abs/2305.05688} {arXiv:2305.05688} \BibitemShut {NoStop}%
\bibitem [{\citenamefont {Bijnens}\ \emph {et~al.}(2019)\citenamefont {Bijnens}, \citenamefont {Kampf},\ and\ \citenamefont {Sj\"o}}]{Bijnens:2019eze}%
  \BibitemOpen
  \bibfield  {author} {\bibinfo {author} {\bibfnamefont {J.}~\bibnamefont {Bijnens}}, \bibinfo {author} {\bibfnamefont {K.}~\bibnamefont {Kampf}}, \ and\ \bibinfo {author} {\bibfnamefont {M.}~\bibnamefont {Sj\"o}},\ }\href {\doibase 10.1007/JHEP11(2019)074} {\bibfield  {journal} {\bibinfo  {journal} {JHEP}\ }\textbf {\bibinfo {volume} {11}},\ \bibinfo {pages} {074} (\bibinfo {year} {2019})},\ \bibinfo {note} {[Erratum: JHEP 03, 066 (2021)]},\ \Eprint {http://arxiv.org/abs/1909.13684} {arXiv:1909.13684} \BibitemShut {NoStop}%
\bibitem [{\citenamefont {Kampf}(2021)}]{Kampf:2021jvf}%
  \BibitemOpen
  \bibfield  {author} {\bibinfo {author} {\bibfnamefont {K.}~\bibnamefont {Kampf}},\ }\href {\doibase 10.1007/JHEP12(2021)140} {\bibfield  {journal} {\bibinfo  {journal} {JHEP}\ }\textbf {\bibinfo {volume} {12}},\ \bibinfo {pages} {140} (\bibinfo {year} {2021})},\ \Eprint {http://arxiv.org/abs/2109.11574} {arXiv:2109.11574} \BibitemShut {NoStop}%
\bibitem [{\citenamefont {Low}\ and\ \citenamefont {Yin}(2019)}]{Low:2019ynd}%
  \BibitemOpen
  \bibfield  {author} {\bibinfo {author} {\bibfnamefont {I.}~\bibnamefont {Low}}\ and\ \bibinfo {author} {\bibfnamefont {Z.}~\bibnamefont {Yin}},\ }\href {\doibase 10.1007/JHEP11(2019)078} {\bibfield  {journal} {\bibinfo  {journal} {JHEP}\ }\textbf {\bibinfo {volume} {11}},\ \bibinfo {pages} {078} (\bibinfo {year} {2019})},\ \Eprint {http://arxiv.org/abs/1904.12859} {arXiv:1904.12859} \BibitemShut {NoStop}%
\bibitem [{\citenamefont {Low}\ \emph {et~al.}(2023)\citenamefont {Low}, \citenamefont {Shu}, \citenamefont {Xiao},\ and\ \citenamefont {Zheng}}]{Low:2022iim}%
  \BibitemOpen
  \bibfield  {author} {\bibinfo {author} {\bibfnamefont {I.}~\bibnamefont {Low}}, \bibinfo {author} {\bibfnamefont {J.}~\bibnamefont {Shu}}, \bibinfo {author} {\bibfnamefont {M.-L.}\ \bibnamefont {Xiao}}, \ and\ \bibinfo {author} {\bibfnamefont {Y.-H.}\ \bibnamefont {Zheng}},\ }\href {\doibase 10.1007/JHEP01(2023)031} {\bibfield  {journal} {\bibinfo  {journal} {JHEP}\ }\textbf {\bibinfo {volume} {01}},\ \bibinfo {pages} {031} (\bibinfo {year} {2023})},\ \Eprint {http://arxiv.org/abs/2209.00198} {arXiv:2209.00198} \BibitemShut {NoStop}%
\bibitem [{\citenamefont {Green}\ \emph {et~al.}(2023)\citenamefont {Green}, \citenamefont {Huang},\ and\ \citenamefont {Shen}}]{Green:2022slj}%
  \BibitemOpen
  \bibfield  {author} {\bibinfo {author} {\bibfnamefont {D.}~\bibnamefont {Green}}, \bibinfo {author} {\bibfnamefont {Y.}~\bibnamefont {Huang}}, \ and\ \bibinfo {author} {\bibfnamefont {C.-H.}\ \bibnamefont {Shen}},\ }\href {\doibase 10.1103/PhysRevD.107.043534} {\bibfield  {journal} {\bibinfo  {journal} {Phys. Rev. D}\ }\textbf {\bibinfo {volume} {107}},\ \bibinfo {pages} {043534} (\bibinfo {year} {2023})},\ \Eprint {http://arxiv.org/abs/2208.14544} {arXiv:2208.14544} \BibitemShut {NoStop}%
\bibitem [{\citenamefont {Cheung}\ and\ \citenamefont {Shen}(2017)}]{Cheung:2016prv}%
  \BibitemOpen
  \bibfield  {author} {\bibinfo {author} {\bibfnamefont {C.}~\bibnamefont {Cheung}}\ and\ \bibinfo {author} {\bibfnamefont {C.-H.}\ \bibnamefont {Shen}},\ }\href {\doibase 10.1103/PhysRevLett.118.121601} {\bibfield  {journal} {\bibinfo  {journal} {Phys. Rev. Lett.}\ }\textbf {\bibinfo {volume} {118}},\ \bibinfo {pages} {121601} (\bibinfo {year} {2017})},\ \Eprint {http://arxiv.org/abs/1612.00868} {arXiv:1612.00868} \BibitemShut {NoStop}%
\bibitem [{\citenamefont {Cheung}\ \emph {et~al.}(2022)\citenamefont {Cheung}, \citenamefont {Helset},\ and\ \citenamefont {Parra-Martinez}}]{Cheung:2022vnd}%
  \BibitemOpen
  \bibfield  {author} {\bibinfo {author} {\bibfnamefont {C.}~\bibnamefont {Cheung}}, \bibinfo {author} {\bibfnamefont {A.}~\bibnamefont {Helset}}, \ and\ \bibinfo {author} {\bibfnamefont {J.}~\bibnamefont {Parra-Martinez}},\ }\href {\doibase 10.1103/PhysRevD.106.045016} {\bibfield  {journal} {\bibinfo  {journal} {Phys. Rev. D}\ }\textbf {\bibinfo {volume} {106}},\ \bibinfo {pages} {045016} (\bibinfo {year} {2022})},\ \Eprint {http://arxiv.org/abs/2202.06972} {arXiv:2202.06972} \BibitemShut {NoStop}%
\bibitem [{\citenamefont {Zhou}\ and\ \citenamefont {Wei}(2023)}]{Zhou:2023quv}%
  \BibitemOpen
  \bibfield  {author} {\bibinfo {author} {\bibfnamefont {K.}~\bibnamefont {Zhou}}\ and\ \bibinfo {author} {\bibfnamefont {F.-S.}\ \bibnamefont {Wei}},\ }\href@noop {} {\  (\bibinfo {year} {2023})},\ \Eprint {http://arxiv.org/abs/2306.09733} {arXiv:2306.09733} \BibitemShut {NoStop}%
\bibitem [{\citenamefont {Bern}\ \emph {et~al.}(2019)\citenamefont {Bern}, \citenamefont {Carrasco}, \citenamefont {Chiodaroli}, \citenamefont {Johansson},\ and\ \citenamefont {Roiban}}]{Bern:2019prr}%
  \BibitemOpen
  \bibfield  {author} {\bibinfo {author} {\bibfnamefont {Z.}~\bibnamefont {Bern}}, \bibinfo {author} {\bibfnamefont {J.~J.}\ \bibnamefont {Carrasco}}, \bibinfo {author} {\bibfnamefont {M.}~\bibnamefont {Chiodaroli}}, \bibinfo {author} {\bibfnamefont {H.}~\bibnamefont {Johansson}}, \ and\ \bibinfo {author} {\bibfnamefont {R.}~\bibnamefont {Roiban}},\ }\href@noop {} {\  (\bibinfo {year} {2019})},\ \Eprint {http://arxiv.org/abs/1909.01358} {arXiv:1909.01358} \BibitemShut {NoStop}%
\bibitem [{\citenamefont {Cachazo}\ \emph {et~al.}(2015)\citenamefont {Cachazo}, \citenamefont {He},\ and\ \citenamefont {Yuan}}]{Cachazo:2014xea}%
  \BibitemOpen
  \bibfield  {author} {\bibinfo {author} {\bibfnamefont {F.}~\bibnamefont {Cachazo}}, \bibinfo {author} {\bibfnamefont {S.}~\bibnamefont {He}}, \ and\ \bibinfo {author} {\bibfnamefont {E.~Y.}\ \bibnamefont {Yuan}},\ }\href {\doibase 10.1007/JHEP07(2015)149} {\bibfield  {journal} {\bibinfo  {journal} {JHEP}\ }\textbf {\bibinfo {volume} {07}},\ \bibinfo {pages} {149} (\bibinfo {year} {2015})},\ \Eprint {http://arxiv.org/abs/1412.3479} {arXiv:1412.3479} \BibitemShut {NoStop}%
\bibitem [{\citenamefont {Casali}\ \emph {et~al.}(2015)\citenamefont {Casali}, \citenamefont {Geyer}, \citenamefont {Mason}, \citenamefont {Monteiro},\ and\ \citenamefont {Roehrig}}]{Casali:2015vta}%
  \BibitemOpen
  \bibfield  {author} {\bibinfo {author} {\bibfnamefont {E.}~\bibnamefont {Casali}}, \bibinfo {author} {\bibfnamefont {Y.}~\bibnamefont {Geyer}}, \bibinfo {author} {\bibfnamefont {L.}~\bibnamefont {Mason}}, \bibinfo {author} {\bibfnamefont {R.}~\bibnamefont {Monteiro}}, \ and\ \bibinfo {author} {\bibfnamefont {K.~A.}\ \bibnamefont {Roehrig}},\ }\href {\doibase 10.1007/JHEP11(2015)038} {\bibfield  {journal} {\bibinfo  {journal} {JHEP}\ }\textbf {\bibinfo {volume} {11}},\ \bibinfo {pages} {038} (\bibinfo {year} {2015})},\ \Eprint {http://arxiv.org/abs/1506.08771} {arXiv:1506.08771} \BibitemShut {NoStop}%
\bibitem [{\citenamefont {Arkani-Hamed}\ \emph {et~al.}(2018{\natexlab{a}})\citenamefont {Arkani-Hamed}, \citenamefont {Rodina},\ and\ \citenamefont {Trnka}}]{Arkani-Hamed:2016rak}%
  \BibitemOpen
  \bibfield  {author} {\bibinfo {author} {\bibfnamefont {N.}~\bibnamefont {Arkani-Hamed}}, \bibinfo {author} {\bibfnamefont {L.}~\bibnamefont {Rodina}}, \ and\ \bibinfo {author} {\bibfnamefont {J.}~\bibnamefont {Trnka}},\ }\href {\doibase 10.1103/PhysRevLett.120.231602} {\bibfield  {journal} {\bibinfo  {journal} {Phys. Rev. Lett.}\ }\textbf {\bibinfo {volume} {120}},\ \bibinfo {pages} {231602} (\bibinfo {year} {2018}{\natexlab{a}})},\ \Eprint {http://arxiv.org/abs/1612.02797} {arXiv:1612.02797} \BibitemShut {NoStop}%
\bibitem [{\citenamefont {Rodina}(2019)}]{Rodina:2016jyz}%
  \BibitemOpen
  \bibfield  {author} {\bibinfo {author} {\bibfnamefont {L.}~\bibnamefont {Rodina}},\ }\href {\doibase 10.1007/JHEP09(2019)084} {\bibfield  {journal} {\bibinfo  {journal} {JHEP}\ }\textbf {\bibinfo {volume} {09}},\ \bibinfo {pages} {084} (\bibinfo {year} {2019})},\ \Eprint {http://arxiv.org/abs/1612.06342} {arXiv:1612.06342} \BibitemShut {NoStop}%
\bibitem [{\citenamefont {Carrasco}\ and\ \citenamefont {Rodina}(2019)}]{Carrasco:2019qwr}%
  \BibitemOpen
  \bibfield  {author} {\bibinfo {author} {\bibfnamefont {J.~J.~M.}\ \bibnamefont {Carrasco}}\ and\ \bibinfo {author} {\bibfnamefont {L.}~\bibnamefont {Rodina}},\ }\href {\doibase 10.1103/PhysRevD.100.125007} {\bibfield  {journal} {\bibinfo  {journal} {Phys. Rev. D}\ }\textbf {\bibinfo {volume} {100}},\ \bibinfo {pages} {125007} (\bibinfo {year} {2019})},\ \Eprint {http://arxiv.org/abs/1908.08033} {arXiv:1908.08033} \BibitemShut {NoStop}%
\bibitem [{\citenamefont {Arkani-Hamed}\ \emph {et~al.}(2023{\natexlab{a}})\citenamefont {Arkani-Hamed}, \citenamefont {Frost}, \citenamefont {Salvatori}, \citenamefont {Plamondon},\ and\ \citenamefont {Thomas}}]{Arkani-Hamed:2023lbd}%
  \BibitemOpen
  \bibfield  {author} {\bibinfo {author} {\bibfnamefont {N.}~\bibnamefont {Arkani-Hamed}}, \bibinfo {author} {\bibfnamefont {H.}~\bibnamefont {Frost}}, \bibinfo {author} {\bibfnamefont {G.}~\bibnamefont {Salvatori}}, \bibinfo {author} {\bibfnamefont {P.-G.}\ \bibnamefont {Plamondon}}, \ and\ \bibinfo {author} {\bibfnamefont {H.}~\bibnamefont {Thomas}},\ }\href@noop {} {\  (\bibinfo {year} {2023}{\natexlab{a}})},\ \Eprint {http://arxiv.org/abs/2309.15913} {arXiv:2309.15913} \BibitemShut {NoStop}%
\bibitem [{\citenamefont {Arkani-Hamed}\ \emph {et~al.}(2023{\natexlab{b}})\citenamefont {Arkani-Hamed}, \citenamefont {Frost}, \citenamefont {Salvatori}, \citenamefont {Plamondon},\ and\ \citenamefont {Thomas}}]{Arkani-Hamed:2023mvg}%
  \BibitemOpen
  \bibfield  {author} {\bibinfo {author} {\bibfnamefont {N.}~\bibnamefont {Arkani-Hamed}}, \bibinfo {author} {\bibfnamefont {H.}~\bibnamefont {Frost}}, \bibinfo {author} {\bibfnamefont {G.}~\bibnamefont {Salvatori}}, \bibinfo {author} {\bibfnamefont {P.-G.}\ \bibnamefont {Plamondon}}, \ and\ \bibinfo {author} {\bibfnamefont {H.}~\bibnamefont {Thomas}},\ }\href@noop {} {\  (\bibinfo {year} {2023}{\natexlab{b}})},\ \Eprint {http://arxiv.org/abs/2311.09284} {arXiv:2311.09284} \BibitemShut {NoStop}%
\bibitem [{\citenamefont {Arkani-Hamed}\ \emph {et~al.}(2023{\natexlab{c}})\citenamefont {Arkani-Hamed}, \citenamefont {Cao}, \citenamefont {Dong}, \citenamefont {Figueiredo},\ and\ \citenamefont {He}}]{Arkani-Hamed:2023swr}%
  \BibitemOpen
  \bibfield  {author} {\bibinfo {author} {\bibfnamefont {N.}~\bibnamefont {Arkani-Hamed}}, \bibinfo {author} {\bibfnamefont {Q.}~\bibnamefont {Cao}}, \bibinfo {author} {\bibfnamefont {J.}~\bibnamefont {Dong}}, \bibinfo {author} {\bibfnamefont {C.}~\bibnamefont {Figueiredo}}, \ and\ \bibinfo {author} {\bibfnamefont {S.}~\bibnamefont {He}},\ }\href@noop {} {\  (\bibinfo {year} {2023}{\natexlab{c}})},\ \Eprint {http://arxiv.org/abs/2312.16282} {arXiv:2312.16282} \BibitemShut {NoStop}%
\bibitem [{\citenamefont {Arkani-Hamed}\ \emph {et~al.}(pear{\natexlab{a}})\citenamefont {Arkani-Hamed}, \citenamefont {Frost},\ and\ \citenamefont {Salvatori}}]{nima}%
  \BibitemOpen
  \bibfield  {author} {\bibinfo {author} {\bibfnamefont {N.}~\bibnamefont {Arkani-Hamed}}, \bibinfo {author} {\bibfnamefont {H.}~\bibnamefont {Frost}}, \ and\ \bibinfo {author} {\bibfnamefont {G.}~\bibnamefont {Salvatori}},\ }\href@noop {} {\  (\bibinfo {year} {{\it to appear}}{\natexlab{a}})}\BibitemShut {NoStop}%
\bibitem [{\citenamefont {Arkani-Hamed}\ \emph {et~al.}(pear{\natexlab{b}})\citenamefont {Arkani-Hamed}, \citenamefont {Cao}, \citenamefont {Dong}, \citenamefont {Figueiredo},\ and\ \citenamefont {He}}]{nima2}%
  \BibitemOpen
  \bibfield  {author} {\bibinfo {author} {\bibfnamefont {N.}~\bibnamefont {Arkani-Hamed}}, \bibinfo {author} {\bibfnamefont {Q.}~\bibnamefont {Cao}}, \bibinfo {author} {\bibfnamefont {J.}~\bibnamefont {Dong}}, \bibinfo {author} {\bibfnamefont {C.}~\bibnamefont {Figueiredo}}, \ and\ \bibinfo {author} {\bibfnamefont {S.}~\bibnamefont {He}},\ }\href@noop {} {\  (\bibinfo {year} {{\it to appear}}{\natexlab{b}})}\BibitemShut {NoStop}%
\bibitem [{\citenamefont {Arkani-Hamed}\ \emph {et~al.}(2011)\citenamefont {Arkani-Hamed}, \citenamefont {Bourjaily}, \citenamefont {Cachazo}, \citenamefont {Caron-Huot},\ and\ \citenamefont {Trnka}}]{Arkani-Hamed:2010zjl}%
  \BibitemOpen
  \bibfield  {author} {\bibinfo {author} {\bibfnamefont {N.}~\bibnamefont {Arkani-Hamed}}, \bibinfo {author} {\bibfnamefont {J.~L.}\ \bibnamefont {Bourjaily}}, \bibinfo {author} {\bibfnamefont {F.}~\bibnamefont {Cachazo}}, \bibinfo {author} {\bibfnamefont {S.}~\bibnamefont {Caron-Huot}}, \ and\ \bibinfo {author} {\bibfnamefont {J.}~\bibnamefont {Trnka}},\ }\href {\doibase 10.1007/JHEP01(2011)041} {\bibfield  {journal} {\bibinfo  {journal} {JHEP}\ }\textbf {\bibinfo {volume} {01}},\ \bibinfo {pages} {041} (\bibinfo {year} {2011})},\ \Eprint {http://arxiv.org/abs/1008.2958} {arXiv:1008.2958} \BibitemShut {NoStop}%
\bibitem [{\citenamefont {Arkani-Hamed}\ \emph {et~al.}(2012)\citenamefont {Arkani-Hamed}, \citenamefont {Bourjaily}, \citenamefont {Cachazo},\ and\ \citenamefont {Trnka}}]{Arkani-Hamed:2010pyv}%
  \BibitemOpen
  \bibfield  {author} {\bibinfo {author} {\bibfnamefont {N.}~\bibnamefont {Arkani-Hamed}}, \bibinfo {author} {\bibfnamefont {J.~L.}\ \bibnamefont {Bourjaily}}, \bibinfo {author} {\bibfnamefont {F.}~\bibnamefont {Cachazo}}, \ and\ \bibinfo {author} {\bibfnamefont {J.}~\bibnamefont {Trnka}},\ }\href {\doibase 10.1007/JHEP06(2012)125} {\bibfield  {journal} {\bibinfo  {journal} {JHEP}\ }\textbf {\bibinfo {volume} {06}},\ \bibinfo {pages} {125} (\bibinfo {year} {2012})},\ \Eprint {http://arxiv.org/abs/1012.6032} {arXiv:1012.6032} \BibitemShut {NoStop}%
\bibitem [{\citenamefont {Bourjaily}\ \emph {et~al.}(2015)\citenamefont {Bourjaily}, \citenamefont {Caron-Huot},\ and\ \citenamefont {Trnka}}]{Bourjaily:2013mma}%
  \BibitemOpen
  \bibfield  {author} {\bibinfo {author} {\bibfnamefont {J.~L.}\ \bibnamefont {Bourjaily}}, \bibinfo {author} {\bibfnamefont {S.}~\bibnamefont {Caron-Huot}}, \ and\ \bibinfo {author} {\bibfnamefont {J.}~\bibnamefont {Trnka}},\ }\href {\doibase 10.1007/JHEP01(2015)001} {\bibfield  {journal} {\bibinfo  {journal} {JHEP}\ }\textbf {\bibinfo {volume} {01}},\ \bibinfo {pages} {001} (\bibinfo {year} {2015})},\ \Eprint {http://arxiv.org/abs/1303.4734} {arXiv:1303.4734} \BibitemShut {NoStop}%
\bibitem [{\citenamefont {Bourjaily}\ and\ \citenamefont {Trnka}(2015)}]{Bourjaily:2015jna}%
  \BibitemOpen
  \bibfield  {author} {\bibinfo {author} {\bibfnamefont {J.~L.}\ \bibnamefont {Bourjaily}}\ and\ \bibinfo {author} {\bibfnamefont {J.}~\bibnamefont {Trnka}},\ }\href {\doibase 10.1007/JHEP08(2015)119} {\bibfield  {journal} {\bibinfo  {journal} {JHEP}\ }\textbf {\bibinfo {volume} {08}},\ \bibinfo {pages} {119} (\bibinfo {year} {2015})},\ \Eprint {http://arxiv.org/abs/1505.05886} {arXiv:1505.05886} \BibitemShut {NoStop}%
\bibitem [{\citenamefont {Drummond}\ \emph {et~al.}(2010{\natexlab{a}})\citenamefont {Drummond}, \citenamefont {Henn}, \citenamefont {Korchemsky},\ and\ \citenamefont {Sokatchev}}]{Drummond:2007au}%
  \BibitemOpen
  \bibfield  {author} {\bibinfo {author} {\bibfnamefont {J.~M.}\ \bibnamefont {Drummond}}, \bibinfo {author} {\bibfnamefont {J.}~\bibnamefont {Henn}}, \bibinfo {author} {\bibfnamefont {G.~P.}\ \bibnamefont {Korchemsky}}, \ and\ \bibinfo {author} {\bibfnamefont {E.}~\bibnamefont {Sokatchev}},\ }\href {\doibase 10.1016/j.nuclphysb.2009.10.013} {\bibfield  {journal} {\bibinfo  {journal} {Nucl. Phys. B}\ }\textbf {\bibinfo {volume} {826}},\ \bibinfo {pages} {337} (\bibinfo {year} {2010}{\natexlab{a}})},\ \Eprint {http://arxiv.org/abs/0712.1223} {arXiv:0712.1223} \BibitemShut {NoStop}%
\bibitem [{\citenamefont {Drummond}\ \emph {et~al.}(2010{\natexlab{b}})\citenamefont {Drummond}, \citenamefont {Henn}, \citenamefont {Korchemsky},\ and\ \citenamefont {Sokatchev}}]{Drummond:2008vq}%
  \BibitemOpen
  \bibfield  {author} {\bibinfo {author} {\bibfnamefont {J.~M.}\ \bibnamefont {Drummond}}, \bibinfo {author} {\bibfnamefont {J.}~\bibnamefont {Henn}}, \bibinfo {author} {\bibfnamefont {G.~P.}\ \bibnamefont {Korchemsky}}, \ and\ \bibinfo {author} {\bibfnamefont {E.}~\bibnamefont {Sokatchev}},\ }\href {\doibase 10.1016/j.nuclphysb.2009.11.022} {\bibfield  {journal} {\bibinfo  {journal} {Nucl. Phys. B}\ }\textbf {\bibinfo {volume} {828}},\ \bibinfo {pages} {317} (\bibinfo {year} {2010}{\natexlab{b}})},\ \Eprint {http://arxiv.org/abs/0807.1095} {arXiv:0807.1095} \BibitemShut {NoStop}%
\bibitem [{\citenamefont {Arkani-Hamed}\ \emph {et~al.}(2016)\citenamefont {Arkani-Hamed}, \citenamefont {Bourjaily}, \citenamefont {Cachazo}, \citenamefont {Goncharov}, \citenamefont {Postnikov},\ and\ \citenamefont {Trnka}}]{Arkani-Hamed:2012zlh}%
  \BibitemOpen
  \bibfield  {author} {\bibinfo {author} {\bibfnamefont {N.}~\bibnamefont {Arkani-Hamed}}, \bibinfo {author} {\bibfnamefont {J.~L.}\ \bibnamefont {Bourjaily}}, \bibinfo {author} {\bibfnamefont {F.}~\bibnamefont {Cachazo}}, \bibinfo {author} {\bibfnamefont {A.~B.}\ \bibnamefont {Goncharov}}, \bibinfo {author} {\bibfnamefont {A.}~\bibnamefont {Postnikov}}, \ and\ \bibinfo {author} {\bibfnamefont {J.}~\bibnamefont {Trnka}},\ }\href {\doibase 10.1017/CBO9781316091548} {}\ (\bibinfo  {publisher} {Cambridge University Press},\ \bibinfo {year} {2016})\ \Eprint {http://arxiv.org/abs/1212.5605} {arXiv:1212.5605} \BibitemShut {NoStop}%
\bibitem [{\citenamefont {Arkani-Hamed}\ and\ \citenamefont {Trnka}(2014)}]{Arkani-Hamed:2013jha}%
  \BibitemOpen
  \bibfield  {author} {\bibinfo {author} {\bibfnamefont {N.}~\bibnamefont {Arkani-Hamed}}\ and\ \bibinfo {author} {\bibfnamefont {J.}~\bibnamefont {Trnka}},\ }\href {\doibase 10.1007/JHEP10(2014)030} {\bibfield  {journal} {\bibinfo  {journal} {JHEP}\ }\textbf {\bibinfo {volume} {10}},\ \bibinfo {pages} {030} (\bibinfo {year} {2014})},\ \Eprint {http://arxiv.org/abs/1312.2007} {arXiv:1312.2007} \BibitemShut {NoStop}%
\bibitem [{\citenamefont {Arkani-Hamed}\ \emph {et~al.}(2018{\natexlab{b}})\citenamefont {Arkani-Hamed}, \citenamefont {Thomas},\ and\ \citenamefont {Trnka}}]{Arkani-Hamed:2017vfh}%
  \BibitemOpen
  \bibfield  {author} {\bibinfo {author} {\bibfnamefont {N.}~\bibnamefont {Arkani-Hamed}}, \bibinfo {author} {\bibfnamefont {H.}~\bibnamefont {Thomas}}, \ and\ \bibinfo {author} {\bibfnamefont {J.}~\bibnamefont {Trnka}},\ }\href {\doibase 10.1007/JHEP01(2018)016} {\bibfield  {journal} {\bibinfo  {journal} {JHEP}\ }\textbf {\bibinfo {volume} {01}},\ \bibinfo {pages} {016} (\bibinfo {year} {2018}{\natexlab{b}})},\ \Eprint {http://arxiv.org/abs/1704.05069} {arXiv:1704.05069} \BibitemShut {NoStop}%
\bibitem [{\citenamefont {Damgaard}\ \emph {et~al.}(2019)\citenamefont {Damgaard}, \citenamefont {Ferro}, \citenamefont {Lukowski},\ and\ \citenamefont {Parisi}}]{Damgaard:2019ztj}%
  \BibitemOpen
  \bibfield  {author} {\bibinfo {author} {\bibfnamefont {D.}~\bibnamefont {Damgaard}}, \bibinfo {author} {\bibfnamefont {L.}~\bibnamefont {Ferro}}, \bibinfo {author} {\bibfnamefont {T.}~\bibnamefont {Lukowski}}, \ and\ \bibinfo {author} {\bibfnamefont {M.}~\bibnamefont {Parisi}},\ }\href {\doibase 10.1007/JHEP08(2019)042} {\bibfield  {journal} {\bibinfo  {journal} {JHEP}\ }\textbf {\bibinfo {volume} {08}},\ \bibinfo {pages} {042} (\bibinfo {year} {2019})},\ \Eprint {http://arxiv.org/abs/1905.04216} {arXiv:1905.04216} \BibitemShut {NoStop}%
\bibitem [{\citenamefont {Ferro}\ and\ \citenamefont {Lukowski}(2023)}]{Ferro:2022abq}%
  \BibitemOpen
  \bibfield  {author} {\bibinfo {author} {\bibfnamefont {L.}~\bibnamefont {Ferro}}\ and\ \bibinfo {author} {\bibfnamefont {T.}~\bibnamefont {Lukowski}},\ }\href {\doibase 10.1007/JHEP05(2023)183} {\bibfield  {journal} {\bibinfo  {journal} {JHEP}\ }\textbf {\bibinfo {volume} {05}},\ \bibinfo {pages} {183} (\bibinfo {year} {2023})},\ \Eprint {http://arxiv.org/abs/2210.01127} {arXiv:2210.01127} \BibitemShut {NoStop}%
\bibitem [{\citenamefont {Even-Zohar}\ \emph {et~al.}(2021)\citenamefont {Even-Zohar}, \citenamefont {Lakrec},\ and\ \citenamefont {Tessler}}]{Even-Zohar:2021sec}%
  \BibitemOpen
  \bibfield  {author} {\bibinfo {author} {\bibfnamefont {C.}~\bibnamefont {Even-Zohar}}, \bibinfo {author} {\bibfnamefont {T.}~\bibnamefont {Lakrec}}, \ and\ \bibinfo {author} {\bibfnamefont {R.~J.}\ \bibnamefont {Tessler}},\ }\href@noop {} {\  (\bibinfo {year} {2021})},\ \Eprint {http://arxiv.org/abs/2112.02703} {arXiv:2112.02703} \BibitemShut {NoStop}%
\bibitem [{\citenamefont {Even-Zohar}\ \emph {et~al.}(2023)\citenamefont {Even-Zohar}, \citenamefont {Lakrec}, \citenamefont {Parisi}, \citenamefont {Tessler}, \citenamefont {Sherman-Bennett},\ and\ \citenamefont {Williams}}]{Even-Zohar:2023del}%
  \BibitemOpen
  \bibfield  {author} {\bibinfo {author} {\bibfnamefont {C.}~\bibnamefont {Even-Zohar}}, \bibinfo {author} {\bibfnamefont {T.}~\bibnamefont {Lakrec}}, \bibinfo {author} {\bibfnamefont {M.}~\bibnamefont {Parisi}}, \bibinfo {author} {\bibfnamefont {R.}~\bibnamefont {Tessler}}, \bibinfo {author} {\bibfnamefont {M.}~\bibnamefont {Sherman-Bennett}}, \ and\ \bibinfo {author} {\bibfnamefont {L.}~\bibnamefont {Williams}},\ }\href@noop {} {\  (\bibinfo {year} {2023})},\ \Eprint {http://arxiv.org/abs/2310.17727} {arXiv:2310.17727} \BibitemShut {NoStop}%
\bibitem [{\citenamefont {Dian}\ \emph {et~al.}(2023)\citenamefont {Dian}, \citenamefont {Heslop},\ and\ \citenamefont {Stewart}}]{Dian:2022tpf}%
  \BibitemOpen
  \bibfield  {author} {\bibinfo {author} {\bibfnamefont {G.}~\bibnamefont {Dian}}, \bibinfo {author} {\bibfnamefont {P.}~\bibnamefont {Heslop}}, \ and\ \bibinfo {author} {\bibfnamefont {A.}~\bibnamefont {Stewart}},\ }\href {\doibase 10.21468/SciPostPhys.15.3.098} {\bibfield  {journal} {\bibinfo  {journal} {SciPost Phys.}\ }\textbf {\bibinfo {volume} {15}},\ \bibinfo {pages} {098} (\bibinfo {year} {2023})},\ \Eprint {http://arxiv.org/abs/2207.12464} {arXiv:2207.12464} \BibitemShut {NoStop}%
\bibitem [{\citenamefont {Arkani-Hamed}\ \emph {et~al.}(2022)\citenamefont {Arkani-Hamed}, \citenamefont {Henn},\ and\ \citenamefont {Trnka}}]{Arkani-Hamed:2021iya}%
  \BibitemOpen
  \bibfield  {author} {\bibinfo {author} {\bibfnamefont {N.}~\bibnamefont {Arkani-Hamed}}, \bibinfo {author} {\bibfnamefont {J.}~\bibnamefont {Henn}}, \ and\ \bibinfo {author} {\bibfnamefont {J.}~\bibnamefont {Trnka}},\ }\href {\doibase 10.1007/JHEP03(2022)108} {\bibfield  {journal} {\bibinfo  {journal} {JHEP}\ }\textbf {\bibinfo {volume} {03}},\ \bibinfo {pages} {108} (\bibinfo {year} {2022})},\ \Eprint {http://arxiv.org/abs/2112.06956} {arXiv:2112.06956} \BibitemShut {NoStop}%
\bibitem [{\citenamefont {Brown}\ \emph {et~al.}(2023{\natexlab{b}})\citenamefont {Brown}, \citenamefont {Oktem}, \citenamefont {Paranjape},\ and\ \citenamefont {Trnka}}]{Brown:2023mqi}%
  \BibitemOpen
  \bibfield  {author} {\bibinfo {author} {\bibfnamefont {T.~V.}\ \bibnamefont {Brown}}, \bibinfo {author} {\bibfnamefont {U.}~\bibnamefont {Oktem}}, \bibinfo {author} {\bibfnamefont {S.}~\bibnamefont {Paranjape}}, \ and\ \bibinfo {author} {\bibfnamefont {J.}~\bibnamefont {Trnka}},\ }\href@noop {} {\  (\bibinfo {year} {2023}{\natexlab{b}})},\ \Eprint {http://arxiv.org/abs/2312.17736} {arXiv:2312.17736} \BibitemShut {NoStop}%
\bibitem [{\citenamefont {Caron-Huot}(2011)}]{Caron-Huot:2010fvq}%
  \BibitemOpen
  \bibfield  {author} {\bibinfo {author} {\bibfnamefont {S.}~\bibnamefont {Caron-Huot}},\ }\href {\doibase 10.1007/JHEP05(2011)080} {\bibfield  {journal} {\bibinfo  {journal} {JHEP}\ }\textbf {\bibinfo {volume} {05}},\ \bibinfo {pages} {080} (\bibinfo {year} {2011})},\ \Eprint {http://arxiv.org/abs/1007.3224} {arXiv:1007.3224} \BibitemShut {NoStop}%
\bibitem [{\citenamefont {Bartsch}\ \emph {et~al.}(2022)\citenamefont {Bartsch}, \citenamefont {Kampf},\ and\ \citenamefont {Trnka}}]{Bartsch:2022pyi}%
  \BibitemOpen
  \bibfield  {author} {\bibinfo {author} {\bibfnamefont {C.}~\bibnamefont {Bartsch}}, \bibinfo {author} {\bibfnamefont {K.}~\bibnamefont {Kampf}}, \ and\ \bibinfo {author} {\bibfnamefont {J.}~\bibnamefont {Trnka}},\ }\href {\doibase 10.1103/PhysRevD.106.076008} {\bibfield  {journal} {\bibinfo  {journal} {Phys. Rev. D}\ }\textbf {\bibinfo {volume} {106}},\ \bibinfo {pages} {076008} (\bibinfo {year} {2022})},\ \Eprint {http://arxiv.org/abs/2206.04694} {arXiv:2206.04694} \BibitemShut {NoStop}%
\bibitem [{\citenamefont {Arkani-Hamed}\ \emph {et~al.}(2018{\natexlab{c}})\citenamefont {Arkani-Hamed}, \citenamefont {Bai}, \citenamefont {He},\ and\ \citenamefont {Yan}}]{Arkani-Hamed:2017mur}%
  \BibitemOpen
  \bibfield  {author} {\bibinfo {author} {\bibfnamefont {N.}~\bibnamefont {Arkani-Hamed}}, \bibinfo {author} {\bibfnamefont {Y.}~\bibnamefont {Bai}}, \bibinfo {author} {\bibfnamefont {S.}~\bibnamefont {He}}, \ and\ \bibinfo {author} {\bibfnamefont {G.}~\bibnamefont {Yan}},\ }\href {\doibase 10.1007/JHEP05(2018)096} {\bibfield  {journal} {\bibinfo  {journal} {JHEP}\ }\textbf {\bibinfo {volume} {05}},\ \bibinfo {pages} {096} (\bibinfo {year} {2018}{\natexlab{c}})},\ \Eprint {http://arxiv.org/abs/1711.09102} {arXiv:1711.09102} \BibitemShut {NoStop}%
\bibitem [{\citenamefont {Cachazo}\ \emph {et~al.}(2016)\citenamefont {Cachazo}, \citenamefont {Cha},\ and\ \citenamefont {Mizera}}]{Cachazo:2016njl}%
  \BibitemOpen
  \bibfield  {author} {\bibinfo {author} {\bibfnamefont {F.}~\bibnamefont {Cachazo}}, \bibinfo {author} {\bibfnamefont {P.}~\bibnamefont {Cha}}, \ and\ \bibinfo {author} {\bibfnamefont {S.}~\bibnamefont {Mizera}},\ }\href {\doibase 10.1007/JHEP06(2016)170} {\bibfield  {journal} {\bibinfo  {journal} {JHEP}\ }\textbf {\bibinfo {volume} {06}},\ \bibinfo {pages} {170} (\bibinfo {year} {2016})},\ \Eprint {http://arxiv.org/abs/1604.03893} {arXiv:1604.03893} \BibitemShut {NoStop}%
\bibitem [{not()}]{noterecursion}%
  \BibitemOpen
  \href@noop {} {}\bibinfo {note} {Let us only briefly mention that a similar form for mixed correlators exists with the NLSM correlators on the right-hand side, multiplied now by inverse $X_{ij}^{-1}$ factors. This closes the mentioned recursion. For details see \cite{inprogress}.}\BibitemShut {Stop}%
\bibitem [{\citenamefont {Bartsch}\ \emph {et~al.}(ress)\citenamefont {Bartsch}, \citenamefont {Kampf}, \citenamefont {Novotny},\ and\ \citenamefont {Trnka}}]{inprogress}%
  \BibitemOpen
  \bibfield  {author} {\bibinfo {author} {\bibfnamefont {C.}~\bibnamefont {Bartsch}}, \bibinfo {author} {\bibfnamefont {K.}~\bibnamefont {Kampf}}, \bibinfo {author} {\bibfnamefont {J.}~\bibnamefont {Novotny}}, \ and\ \bibinfo {author} {\bibfnamefont {J.}~\bibnamefont {Trnka}},\ }\href@noop {} {\  (\bibinfo {year} {{\it in progress}})}\BibitemShut {NoStop}%
\bibitem [{\citenamefont {Hinterbichler}\ and\ \citenamefont {Joyce}(2015)}]{Hinterbichler:2015pqa}%
  \BibitemOpen
  \bibfield  {author} {\bibinfo {author} {\bibfnamefont {K.}~\bibnamefont {Hinterbichler}}\ and\ \bibinfo {author} {\bibfnamefont {A.}~\bibnamefont {Joyce}},\ }\href {\doibase 10.1103/PhysRevD.92.023503} {\bibfield  {journal} {\bibinfo  {journal} {Phys. Rev. D}\ }\textbf {\bibinfo {volume} {92}},\ \bibinfo {pages} {023503} (\bibinfo {year} {2015})},\ \Eprint {http://arxiv.org/abs/1501.07600} {arXiv:1501.07600} \BibitemShut {NoStop}%
\bibitem [{\citenamefont {Novotny}(2017)}]{Novotny:2016jkh}%
  \BibitemOpen
  \bibfield  {author} {\bibinfo {author} {\bibfnamefont {J.}~\bibnamefont {Novotny}},\ }\href {\doibase 10.1103/PhysRevD.95.065019} {\bibfield  {journal} {\bibinfo  {journal} {Phys. Rev. D}\ }\textbf {\bibinfo {volume} {95}},\ \bibinfo {pages} {065019} (\bibinfo {year} {2017})},\ \Eprint {http://arxiv.org/abs/1612.01738} {arXiv:1612.01738} \BibitemShut {NoStop}%
\bibitem [{\citenamefont {Cheung}\ \emph {et~al.}(2018)\citenamefont {Cheung}, \citenamefont {Kampf}, \citenamefont {Novotny}, \citenamefont {Shen}, \citenamefont {Trnka},\ and\ \citenamefont {Wen}}]{Cheung:2018oki}%
  \BibitemOpen
  \bibfield  {author} {\bibinfo {author} {\bibfnamefont {C.}~\bibnamefont {Cheung}}, \bibinfo {author} {\bibfnamefont {K.}~\bibnamefont {Kampf}}, \bibinfo {author} {\bibfnamefont {J.}~\bibnamefont {Novotny}}, \bibinfo {author} {\bibfnamefont {C.-H.}\ \bibnamefont {Shen}}, \bibinfo {author} {\bibfnamefont {J.}~\bibnamefont {Trnka}}, \ and\ \bibinfo {author} {\bibfnamefont {C.}~\bibnamefont {Wen}},\ }\href {\doibase 10.1103/PhysRevLett.120.261602} {\bibfield  {journal} {\bibinfo  {journal} {Phys. Rev. Lett.}\ }\textbf {\bibinfo {volume} {120}},\ \bibinfo {pages} {261602} (\bibinfo {year} {2018})},\ \Eprint {http://arxiv.org/abs/1801.01496} {arXiv:1801.01496} \BibitemShut {NoStop}%
\bibitem [{\citenamefont {Elvang}\ \emph {et~al.}(2019)\citenamefont {Elvang}, \citenamefont {Hadjiantonis}, \citenamefont {Jones},\ and\ \citenamefont {Paranjape}}]{Elvang:2018dco}%
  \BibitemOpen
  \bibfield  {author} {\bibinfo {author} {\bibfnamefont {H.}~\bibnamefont {Elvang}}, \bibinfo {author} {\bibfnamefont {M.}~\bibnamefont {Hadjiantonis}}, \bibinfo {author} {\bibfnamefont {C.~R.~T.}\ \bibnamefont {Jones}}, \ and\ \bibinfo {author} {\bibfnamefont {S.}~\bibnamefont {Paranjape}},\ }\href {\doibase 10.1007/JHEP01(2019)195} {\bibfield  {journal} {\bibinfo  {journal} {JHEP}\ }\textbf {\bibinfo {volume} {01}},\ \bibinfo {pages} {195} (\bibinfo {year} {2019})},\ \Eprint {http://arxiv.org/abs/1806.06079} {arXiv:1806.06079} \BibitemShut {NoStop}%
\bibitem [{\citenamefont {Roest}\ \emph {et~al.}(2019)\citenamefont {Roest}, \citenamefont {Stefanyszyn},\ and\ \citenamefont {Werkman}}]{Roest:2019oiw}%
  \BibitemOpen
  \bibfield  {author} {\bibinfo {author} {\bibfnamefont {D.}~\bibnamefont {Roest}}, \bibinfo {author} {\bibfnamefont {D.}~\bibnamefont {Stefanyszyn}}, \ and\ \bibinfo {author} {\bibfnamefont {P.}~\bibnamefont {Werkman}},\ }\href {\doibase 10.1007/JHEP08(2019)081} {\bibfield  {journal} {\bibinfo  {journal} {JHEP}\ }\textbf {\bibinfo {volume} {08}},\ \bibinfo {pages} {081} (\bibinfo {year} {2019})},\ \Eprint {http://arxiv.org/abs/1903.08222} {arXiv:1903.08222} \BibitemShut {NoStop}%
\bibitem [{\citenamefont {P\v{r}eu\v{c}il}\ and\ \citenamefont {Novotn\'y}(2019)}]{Preucil:2019nxt}%
  \BibitemOpen
  \bibfield  {author} {\bibinfo {author} {\bibfnamefont {F.}~\bibnamefont {P\v{r}eu\v{c}il}}\ and\ \bibinfo {author} {\bibfnamefont {J.}~\bibnamefont {Novotn\'y}},\ }\href {\doibase 10.1007/JHEP11(2019)166} {\bibfield  {journal} {\bibinfo  {journal} {JHEP}\ }\textbf {\bibinfo {volume} {11}},\ \bibinfo {pages} {166} (\bibinfo {year} {2019})},\ \Eprint {http://arxiv.org/abs/1909.06214} {arXiv:1909.06214} \BibitemShut {NoStop}%
\bibitem [{\citenamefont {Kampf}\ and\ \citenamefont {Novotn\'y}(2020)}]{Kampf:2020tne}%
  \BibitemOpen
  \bibfield  {author} {\bibinfo {author} {\bibfnamefont {K.}~\bibnamefont {Kampf}}\ and\ \bibinfo {author} {\bibfnamefont {J.}~\bibnamefont {Novotn\'y}},\ }\href {\doibase 10.1007/JHEP12(2020)056} {\bibfield  {journal} {\bibinfo  {journal} {JHEP}\ }\textbf {\bibinfo {volume} {12}},\ \bibinfo {pages} {056} (\bibinfo {year} {2020})},\ \Eprint {http://arxiv.org/abs/2009.07940} {arXiv:2009.07940} \BibitemShut {NoStop}%
\bibitem [{\citenamefont {Brauner}(2021)}]{Brauner:2020ezm}%
  \BibitemOpen
  \bibfield  {author} {\bibinfo {author} {\bibfnamefont {T.}~\bibnamefont {Brauner}},\ }\href {\doibase 10.1007/JHEP02(2021)218} {\bibfield  {journal} {\bibinfo  {journal} {JHEP}\ }\textbf {\bibinfo {volume} {02}},\ \bibinfo {pages} {218} (\bibinfo {year} {2021})},\ \Eprint {http://arxiv.org/abs/2008.12078} {arXiv:2008.12078} \BibitemShut {NoStop}%
\bibitem [{\citenamefont {Kampf}\ \emph {et~al.}(2021)\citenamefont {Kampf}, \citenamefont {Novotny}, \citenamefont {Preucil},\ and\ \citenamefont {Trnka}}]{Kampf:2021bet}%
  \BibitemOpen
  \bibfield  {author} {\bibinfo {author} {\bibfnamefont {K.}~\bibnamefont {Kampf}}, \bibinfo {author} {\bibfnamefont {J.}~\bibnamefont {Novotny}}, \bibinfo {author} {\bibfnamefont {F.}~\bibnamefont {Preucil}}, \ and\ \bibinfo {author} {\bibfnamefont {J.}~\bibnamefont {Trnka}},\ }\href {\doibase 10.1007/JHEP07(2021)153} {\bibfield  {journal} {\bibinfo  {journal} {JHEP}\ }\textbf {\bibinfo {volume} {07}},\ \bibinfo {pages} {153} (\bibinfo {year} {2021})},\ \Eprint {http://arxiv.org/abs/2104.10693} {arXiv:2104.10693} \BibitemShut {NoStop}%
\bibitem [{\citenamefont {de~Neeling}\ \emph {et~al.}(2022)\citenamefont {de~Neeling}, \citenamefont {Roest},\ and\ \citenamefont {Veldmeijer}}]{deNeeling:2022tsu}%
  \BibitemOpen
  \bibfield  {author} {\bibinfo {author} {\bibfnamefont {D.}~\bibnamefont {de~Neeling}}, \bibinfo {author} {\bibfnamefont {D.}~\bibnamefont {Roest}}, \ and\ \bibinfo {author} {\bibfnamefont {S.}~\bibnamefont {Veldmeijer}},\ }\href {\doibase 10.1007/JHEP10(2022)066} {\bibfield  {journal} {\bibinfo  {journal} {JHEP}\ }\textbf {\bibinfo {volume} {10}},\ \bibinfo {pages} {066} (\bibinfo {year} {2022})},\ \Eprint {http://arxiv.org/abs/2204.11629} {arXiv:2204.11629} \BibitemShut {NoStop}%
\end{thebibliography}%
